\documentclass[aps,prd,reprint,unsortedaddress,showpacs]{revtex4-1}

\usepackage[utf8]{inputenc}
\usepackage{amssymb,amsmath,textcomp}
\usepackage{ctable}
\usepackage{graphicx}
\usepackage{subfigure}
\usepackage{verbatim}

\begin{document}


\title{Systematic study of actinide and pre-actinide fission modes}

\author{E. Andrade-II}
\affiliation{Instituto de F\'isica, Universidade de S\~ao Paulo, P. O. Box 66318, 05389-970 S\~ao Paulo, SP, Brazil}
\author{G. S. Karapetyan}
\affiliation{Instituto de Fisica, Universidade de S\~ao Paulo, P. O. Box 66318, 05389-970 S\~ao Paulo, SP, Brazil}
\author{A. Deppman}
\affiliation{Instituto de Fisica, Universidade de Sao Paulo, P. O. Box 66318, 05315-970 S\~ao Paulo, SP, Brazil}
\author{Jose L. Bernal-Castillo}
\affiliation{Instituto de Fisica, Universidade de Sao Paulo, P. O. Box 66318, 05389-970 S\~ao Paulo, SP, Brazil}
\author{A. R. Balabekyan}
\affiliation{Yerevan State University, Faculty of Physics, Alex Manoogian 1, Yerevan 0025, Armenia}
\author{N. A. Demekhina}
\affiliation{Yerevan Physics Institute, Alikhanyan Brothers 2, Yerevan 0036, Armenia\\Joint Institute for Nuclear Research 
(JINR), Flerov Laboratory of Nuclear Reactions (LNR), Joliot-Curie 6, Dubna 141980, Moscow region Russia}
\author{J. Adam}
\affiliation{Joint Institute for Nuclear Research (JINR), Djelepov Laboratory of Nuclear Problem (DLNP),
Joliot-Curie 6, Dubna 141980, Moscow region Russia}
\author{F. Garcia}
\affiliation{Universidade Estadual de Santa Cruz, 45662-900 Ilh\'eus, BA, Brazil}
\author{F. Guzm\'an}
\affiliation{Instituto Superior de Tecnologías y Ciencias Aplicadas (InSTEC), Havana, Cuba}
\date{\today}

\begin{abstract}
In this work, we present new experimental data on mass distribution of fission fragments from $^{241}$Am proton-induced fission at $660$ MeV measured at 
the LNR Phasotron (JINR). The systematic analysis of several measured fragment mass distributions from different fission reactions available in the 
literature is also presented. The proton-induced fission of $^{241}$Am, $^{237}$Np and $^{238}$U at 26.5, 62.9 and 660 MeV was studied. The proton-induced fission 
of $^{232}$Th was studied at 26.5, 62.9 and 190 MeV. The fission of $^{208}$Pb also by a proton was investigated at 190, 500 and 1000 MeV. The fission of $^{197}$Au was 
studied for 190 and 800 MeV protons. Bremsstrahlung reactions with maximum photon energies of 50 and 3500 MeV were studied for $^{232}$Th and $^{238}$U. 
The framework of the Random Neck Rupture Model was applied in the analysis. The roles of the neutron excess and of the so called fissility parameter were 
also investigated.
\end{abstract}



\maketitle

\section{Introduction}

The importance of the nuclear fission can hardly be exaggerated. As a matter of technology, the fission of the nucleus is a protagonist
in the production of isotopes for both medical and industrial purposes \cite{nupecc2014}. Such applications are increasingly important to society today. 
Also, it seems the role of nuclear fission is meant to grow regarding power generation since the world need 
for energy has never stopped increasing. Conventional reactors are likely to grow in number and, although the design is continually advancing, 
better solutions for nuclear waste are mandatory. Nowadays, accelerator-driven systems (ADS) are promising due to their capability of 
incinerating nuclear material originated in conventional reactors. Besides, these sub-critical reactors can also produce energy 
\cite{Gudowsky1999, DENERAC2002, Aliberti2004}. Although the neutrons necessary to maintain the processes in the reactor come from spallation 
reactions in the accelerator, fission might occur inside both the target (inside accelerator) and the reactor core \cite{Duijvestijn2001}. 
Therefore a better knowledge of spallation and fission dynamics is mandatory for the developments of such technology.

In spite of so many relevant applications, with some of them successfully at use already, nuclear fission remains a subject for intense
theoretical study because many aspects were not yet elucidated. The concept of multi-modal fission was suggested in several publications 
\cite{Turkevich1951,Ford1960,Pashkevich1971,Wilkins1976,Brosa1990}, where a physical 
meaning to the multi-modal fission hypothesis was provided by arguing that the contribution of shell effects in the liquid drop effectively 
gives rise to valleys on the potential energy surface of the fissioning system. Each valley corresponds to a specific ensemble of deformation 
configurations which leads to a particular fission mode \cite{Mulgin2009}. In this sense, shell effects are thought
to be more relevant at low excitation energy, when one symmetrical fission mode (Superlong) along with at least two asymmetrical ones (Standard 1 and 2)
should be reasonably identifiable. With increasing energy, the nuclear structure relevance is expected to decrease and the symmetrical mode would become
more prominent or even the only one present. However, this is not always the case as neutron-deficient actinides may undergo symmetrical fission already
at low energies and sub-actinides will prefer this fission mode in a wide range of excitation energy \cite{Duijvestijn1999,Duijvestijn2001}. Moreover, recent analysis 
have shown that even at relatively high excitation energy the contribution of asymmetric modes is still present in heavy nuclei fission  \cite{Karapetyan2009,Balabekyan2010,Deppman2013b,Deppman2013c}.

Another issue arises at high energies because while first-chance fission is more likely at low energy, multi-chance fission begins to
play a more important role with increasing excitation. With this energy availability, the fissioning nucleus may emit nucleons several times before 
fission. A great amount of intermediate mass fissioning systems are formed each one with different fission properties and all of them contributing 
to the final fission observables. The relation between the excitation energy and the distribution of fissioning systems was already studied for some 
reactions in another work by Andrade-II et al \cite{Andrade2011} and the conclusion was that the correct description of fission at high energies 
must take into account the distribution of fissioning systems. 

In this paper we report experimental results on fission products from the super-asymmetric fission of $^{241}$Am by 660 MeV protons. Although there is a clear 
correlation between increasing excitation energy and increasing symmetric fission, it seems the precise mechanism that leads to symmetry or asymmetry is yet to be 
determined, considering the examples of exception presented above in the second paragraph of this Introduction. For this reason, a systematic analysis of 
fission fragment production is performed by examining the mass distributions of residual nuclei and fissioning systems, the average neutron excess and the so called fissility parameter, defined as $Z^2/A$. Chung et al 
\cite{Chung1981, Chung1982} developed a criterion based on the fissility parameter in order to determine the probability of symmetric and asymmetric fission following the assumption that nuclear structure does affest fission. Also in this paper 
we present and examine the multi-modal fission model parameters as determined for each reaction studied. Finally, a process called cold cluster emission is 
proposed to explain a specific feature observed in the fragment mass distributions of the pre-actinide nuclei.  

The present paper is organized as follows: The CRISP model, central instrument of this systematic analysis, is presented in Section \ref{sec:Model}. The new 
experimental data, Section \ref{sec:ExpRes}. A general analysis of the fission fragments distributions is presented in Section \ref{sec:FragDistAna}. Section 
\ref{sec:SysAnaFissMod} examines the parameters obtained for each fission mode in this systematics while Section \ref{sec:NucStrucInf} performs a more detailed investigation of how nuclear structure might influence fission. Section \ref{sec:ColdClus} investigate the so called cluster emission hypothesis. Finally, the conclusions are reported in Section \ref{sec:Conc}.


\section{Model} \label{sec:Model}

The CRISP model (acronym for Collaboration Rio - Ilh\'eus - S\~ao Paulo) was designed to calculate nuclear reactions \cite{Deppman2004} in a two step
process. Firstly, the intranuclear cascade begins after a primary interaction with an incident particle \cite{Goncalves1997, dePina1998, Deppman2002a, Deppman2002b}. 
The proton can interact elastically or inelastically producing Delta resonances. A photon can interact according to the channels that go from the 
quasi-deuteron absorption mechanism at around 50 MeV up to hadronization and vector meson production \cite{Deppman2006,Israel2014,Andrade2015}. 
The most relevant nucleonic resonances at intermediate energy are also produced.

Secondary particles, created during the cascade, may interact with other particles or reach the surface of the nucleus constructed as a Fermi gas. 
Particles with kinetic energy above the nuclear potential may leave, otherwise they are reflected. The multicollisional approach allows a more realistic 
simulation of the intranuclear cascade \cite{Deppman2004} since all particles move simultaneously, the order of events being
established by the time sequence between collisions of each pair of particles and respective cross sections. Such an approach makes it natural to verify dynamical aspects, e.g., nuclear density modification and level occupation evolution of the gas. 

The Pauli blocking mechanism is another important feature in the model. Once all nucleons are located in Fermi gas levels and they are allowed to move simultaneously, 
exchanging energy and momentum, the exclusion principle can be verified strictly \cite{Deppman2004}.

All these characteristics together make possible an energetic stop criterion. When no particle has kinetic energy above the potential, mass and atomic number 
and excitation energy can no longer change, the nucleus is considered thermalized and the cascade is finished since the next step regarding the competition 
between fission and evaporation of particles is modeled only statistically. In this second stage, the emission widths are determined according to Weisskopf's model 
\cite{Weisskopf1937} and calculated all relative to the neutron width as

\begin{align}
 \dfrac{\Gamma_p}{\Gamma_n} = \dfrac{E_p}{E_n} \exp \left\lbrace 2 \left[ (a_p E_p)^{1/2} - (a_n E_n)^{1/2} \right] \right\rbrace ,
 \label{gamap}
\end{align}
\noindent for proton emission and
\begin{equation}
 \dfrac{\Gamma_{\alpha}}{\Gamma_n} = \dfrac{2 E_{\alpha}}{E_n} \exp \left\lbrace 2 \left[ (a_{\alpha} E_{\alpha})^{1/2} - (a_n E_n)^{1/2} \right] \right\rbrace ,
 \label{gamaa}  
\end{equation}
\noindent for alpha particles emission with the energy of the possible final states given by 
\begin{align}
 \begin{split}
   E_n &= E - B_n, \\
   E_p &= E - B_p - V_p, \\
   E_{\alpha} &= E - B_{\alpha} - V_{\alpha},
 \end{split}
\label{enerEsta}
\end{align}

\noindent where $E$ is the current energy of the nucleus, $B_n$, $B_p$ and $B_{\alpha}$ are the separation energies for neutrons, protons and alpha particles, 
respectively. $V_p$ and $V_{\alpha}$ are the Coulomb potential for protons and alpha particles.

The density level parameters $a_n$, $a_p$ and $a_{\alpha}$ for neutrons, protons and alpha particles are determined by Dostrovsky's equations \cite{Dostrovsky1958},
\begin{align}
 \begin{split}
  a_n &= \frac{A}{a_1}\left( 1 - a_2\frac{A-2Z}{A^2}\right)^2, \\
  a_p &= \frac{A}{a_3}\left( 1 + a_4\frac{A-2Z}{A^2}\right)^2, \\
  a_{\alpha} &= \frac{A}{a_5}\left( 1 - \frac{a_6}{Z}\right)^2.
 \end{split}
\label{densNiveis}
\end{align}

The fission process follows Bohr and Wheeler model \cite{BohrWheeler1939} with the fission width calculated according to Vandenbosch and Huizenga 
\cite{Vandenbosch1973}, 

\begin{align}
 \dfrac{\Gamma_f}{\Gamma_n} = K_f \exp \left\lbrace 2 \left[ (a_f E_f)^{1/2} - (a_n E_n)^{1/2} \right]  \right\rbrace ,
 \label{gamaf}
\end{align}
\noindent with
\begin{align}
 K_f = K_0 a_n \dfrac{\left[ 2(a_f E_f)^{1/2} - 1 \right]}{(4A^{2/3}a_f E_n)} ,
 \label{Kf}
\end{align}
and,
\begin{align}
 \begin{split}
    E_f &= E - B_f , \\
    a_f &= r_f a_n ,
 \end{split}
 \label{Efaf}
\end{align}
\noindent where $B_f$ is the fission barrier calculated according to Nix model \cite{Nix1972}. $a_f$ is the fission density level parameter with $r_f$ being 
a adjustable parameter.

In case of fission, CRISP model determines the masses of the fission fragments according to the multi-modal fission model, best known in the literature as 
the Random Neck Rupture Model \cite{Pashkevich1971,Brosa1990}. Following its prescription, the fragments are calculated so that each one falls over a Gaussian. 
The Superlong mode (SL) requires only one. Standard 1 (S1), 2 (S2) and the super-asymmetric Standard 3 (S3) require two Gaussians each, one for the heavy fragment 
and the other for the lighter. The positions of the Gaussians, which are the fragment most probable masses for each mode, the widths and normalization constant are
parameters only determined through comparison to the total experimental mass distribution. The charge distribution is also a Gaussian. The total yield for a fragment 
with mass number A and atomic number Z is determined by

\begin{align} 
\begin{split}
 \sigma(A,Z) =& \bigg\{ \sum_i\bigg[ \frac{K_i}{\sqrt{2 \pi} \Gamma_i} \exp\left( -\frac{(A-A^{\rm L}_i)^2}{2 (\Gamma_i)^2}\right) \\
 & + \frac{K_i}{\sqrt{2 \pi}\Gamma_i}\exp\left( -\frac{(A-A^{\rm H}_i)^2}{2(\Gamma_i)^2}\right) \bigg] \\
 & + \frac{K_{\rm SL}}{\sqrt{2 \pi} \Gamma_{\rm SL}} \exp\left( -\frac{(A-A_{\rm SL})^2}{2 (\Gamma_{\rm SL})^2}\right) \bigg\} \\
 & \times \frac{1}{\sqrt{2\pi}\Gamma_{\rm Z}}\exp\left( -\frac{(Z-\overline{Z_0})^2}{2 \Gamma_{\rm Z}^2}\right) 
\label{eqYield}
\end{split}
\end{align}

\noindent where the summation runs over the asymmetric modes. The parameters for the symmetric mode are $K_{\rm SL}$, $A_{\rm SL}$ and $\Gamma_{\rm SL}$, 
while $K_i$ and $\Gamma_i$ are the parameters for the fragments produced in the asymmetric mode $i = \rm S1, S2, S3$. The position parameters $A_i^{{\rm H}({\rm L})}$ 
for the heavy (light) fragments are determined as $A_i^{\rm H} = A_{\rm SL} + D_i$ and $A_i^{\rm L} = A_{\rm SL} - D_i$. The shift $D_i$ is the adjustable parameter.

It has been acknowledged in the literature the possibility of non-symmetrical mass distributions with respect to the average fragment mass, with 
propositions, motivated by experiment, such as taking different contributions from the Gaussians that refer to the heavier and 
the lighter fragment \cite{Duijvestijn1999, Karapetyan2009, Demekhina2010}. Regarding the fitting procedure of the final experimental distributions by  Gaussians, this approach may be the best choice in order to find the optimal agreement with experimental data. However, as far as the fragments 
at the scission point are concerned, different contributions from heavy and light fragments should not be accepted since this would contradict the concept of binary fission. CRISP model realistically follows the history of many targets from primary interaction with 
the projectile up to fission and spallation considering all stages in between. Since the multi-modal fission approach takes place 
at the instant of nuclear break-up our parameters for heavy and light fragments are identical. Any distortion in mass distribution must 
come from the different evaporation chains of the different fragments produced.

According to the Monte Carlo method applied in this study, each simulated fission history leads to a particular fissioning system with a particular $A_{\rm SL}$. This variation of $A_{\rm SL}$ is considered by the CRISP model in Equation \ref{eqYield} making the choice of the fragments unique and folding together the 
fissioning system mass distribution and the fragment mass distribution, what is quite in accordance with reality. 

For the atomic number distribution the parametrization used is \cite{Kudo1998}

\begin{align}
 \overline{Z_0}=\mu_1+\mu_2 A
 \label{eqProbZ}
\end{align}

\noindent for the most probable atomic number of the fragment, and 

\begin{align}
 \Gamma_{\rm Z}=\nu_1+\nu_2 A
 \label{eqWidthZ}
\end{align}

\noindent for the width of the atomic number distribution. $\mu_1$, $\mu_2$, $\nu_1$ and $\nu_2$ are fitting parameters. In the discussion below the dependence 
of distributions on the atomic number Z will not be relevant.

Normalization constant, position and width parameters for each mode in equation (\ref{eqYield}) and $\mu_1$, $\mu_2$, $\nu_1$ and $\nu_2$ in (\ref{eqProbZ}) 
and (\ref{eqWidthZ}) are usually considered free parameters for fitting procedure.

After determination of the fission fragments they are allowed to evaporate following the already mentioned statistical evaporation 
model of Weisskopf. The final products of fission can then be compared to experimental data.

\section{Experimental Results} \label{sec:ExpRes}

Besides the data from other sources that we analyze in this work we also present new data on mass distribution of fission fragments from the reaction 
p(660 MeV) + $^{241}$Am. 

All experimental data in Refs. \cite{Karapetyan2009,Balabekyan2010} represent the cross sections of the elements which have been measured experimentally. 
Although a large number of cross sections have been determined, the data represent only a fraction of the total
isobaric yields. The cross section of a particular isotope may be independent or partly or completely cumulative, depending on decay chains 
of precursors. The beta-decay feeding correction factors for cumulative yield isobaric members can be calculated once the centroid and width of 
the Gaussian are known. In order to obtain the mass-yield distribution, it is necessary to make an estimation of the cross sections for unmeasured 
products. Thus, we have made an assumption of Gaussian charge distribution that is, the independent yield cross section  can be represented by a 
Gaussian curve and applied least-squares method in order to obtain the total isobaric cross sections. So, the solid line on Figs. in 
Refs. \cite{Karapetyan2009,Balabekyan2010} represents the mass-yield distributions obtained in this manner based on the fitting procedure which 
gives the total isobaric cross sections.

All points on Figs. in \cite{Deppman2013b,Deppman2013c} are the total isobaric cross sections from \cite{Karapetyan2009,Balabekyan2010} and the solid 
lines are the calculations which were done by CRISP model in order to compare with experimental mass-yield distribution. We should stress also that in 
\cite{Deppman2013c} we have presented the new data for super-asymmetric fission mode just for $^{238}$U and $^{237}$Np targets for proton-induced fission at 
660 MeV. In the present article we complete the database with the super-asymmetric elements also from $^{241}$Am target.

The number of intermediate mass fragments (IMFs) from $^{241}$Am target in the mass range $28 < A < 69$ and the cross sections of heavy elements in 
the mass range $150<A<191$ from $^{241}$Am, $^{238}$U and $^{237}$Np targets have been obtained in off-line experiment using 
induced-activity method \cite{Karapetyan2015}. These correspond to the new measured data 
that are further analyzed to take into account the cumulative contribution from decay nuclides \cite{Firestone1998}.

%

Tables \ref{tabIMFAm} and \ref{tabfragAmNpU} present the experimentally measured cross sections. The labels C and I correspond to independent and cumulative cross sections, 
respectively.

\begin{table*}
\centering
\caption{New measured IMFs cross sections for the fission of $^{241}$Am by 660 MeV protons. Labels C and I correspond to independent and cumulative cross sections, 
respectively.}
\label{tabIMFAm}
\begin{ruledtabular}
\begin{tabular}{l c c c|c l c c}
Element&Type& Cross section, mb  &&& Element&Type& Cross section, mb \\\cline{1-3} \cline{6-8}
&&&&&&& \\
$^{28}$Mg   & C & 0.27$\pm$0.03 &&& $^{52m}$Mn  & I & 1.2$\pm$0.12\\ 
$^{34m}$Cl  & I & 0.3$\pm$0.04 &&& $^{52}$Fe  & I & 0.11$\pm$0.02  \\ 
$^{38}$S & I & 0.02$\pm$0.004 &&& $^{55}$Co & C & 0.15$\pm$0.02 \\ 
$^{38}$Cl   & I & 0.15$\pm$0.02 &&& $^{56}$Co & I & 0.52$\pm$0.05 \\ 
$^{39}$Cl   & C & 0.016$\pm$0.003   &&& $^{56}$Ni & I & $\leq$0.02 \\ 
$^{41}$Ar   & C & 0.3$\pm$0.04 &&& $^{57}$Co  & I & 0.72$\pm$0.07  \\ 
$^{42}$K    & C & 0.28$\pm$0.05 &&& $^{57}$Ni   & I & 0.057$\pm$0.006 \\ 
$^{43}$K    & C & 0.15$\pm$0.02 &&& $^{58(m+g)}$Co & I & 0.8$\pm$0.08 \\ 
$^{43}$Sc   & C & 0.2$\pm$0.03 &&& $^{59}$Fe  & C & 1.5$\pm$0.15  \\ 
$^{44}$Ar   & I & $\leq$6.2E-3  &&& $^{60(m+g)}$Co & I & 1.82$\pm$0.2  \\ 
$^{44}$K    & I & 0.15$\pm$0.03 &&&  $^{60}$Cu  & C & $\leq$0.065 \\ 
$^{44g}$Sc  & I & 0.14$\pm$0.02  &&&  $^{61}$Co & C & 0.27$\pm$0.03 \\ 
$^{44m}$Sc  & I & 0.09$\pm$0.02 &&& $^{61}$Cu & C & 0.16$\pm$0.02 \\ 
$^{45}$K    & C & 0.07$\pm$0.01 &&& $^{65}$Ni & I & 0.045$\pm$0.006  \\ 
$^{46(m+g)}$Sc&I& 0.53$\pm$0.05 &&& $^{65}$Zn & I & 1.4$\pm$0.14\\ 
$^{47}$Ca   & I & 0.04$\pm$0.008  &&& $^{65}$Ga & C & 0.5$\pm$0.05  \\ 
$^{47}$Sc   & I & 0.4$\pm$0.04    &&&  $^{66}$Ni & I & 0.06$\pm$0.001 \\ 
$^{48}$Sc   & I & 1.11$\pm$0.1 &&& $^{66}$Ga & I & 0.4$\pm$0.04 \\ 
$^{48}$V    & I & 2.9$\pm$0.3 &&& $^{66}$Ge & I & 0.02$\pm$0.004 \\ 
$^{48}$Cr   & I & 0.45$\pm$0.05  &&& $^{67}$Cu & C & 0.5$\pm$0.05  \\ 
$^{49}$Cr   & C & 0.13$\pm$0.02 &&& $^{67}$Ga & C & 0.56$\pm$0.06 \\ 
$^{51}$Cr   & C & 0.35$\pm$0.04 &&& $^{69m}$Zn & I & 1.4$\pm$0.14 \\ 
$^{52g}$Mn  & C & 0.5$\pm$0.05 &&& $^{69}$Ge & C & 0.082$\pm$0.01 \\ 
\end{tabular}
\end{ruledtabular}
\end{table*}

\begin{table*}
\centering
\caption{Measured heavy products cross sections for the fission of $^{237}$Np \cite{Karapetyan2009,Deppman2013c}, $^{238}$U \cite{Balabekyan2010,Deppman2013c} 
and $^{241}$Am by 660 MeV protons. Labels C and I correspond to independent and cumulative cross sections, 
respectively.}
\label{tabfragAmNpU}
\begin{ruledtabular}
\begin{tabular}{c c c c c | c c c c c}
Element&Type& \multicolumn{3}{c|}{Cross section, mb }&Element&Type& \multicolumn{3}{c}{Cross section, mb } \\
& &  &  & (This work) & & &  &  & (This work) \\
& & $^{237}$Np & $^{238}$U & $^{241}$Am & & &$^{237}$Np & $^{238}$U & $^{241}$Am \\  \hline
&&&&&&&&&\\
$^{150}$Pm{}&{}I{}&0.7$\pm$0.07&0.72$\pm$0.07&0.72$\pm$0.07&$^{172}$Lu{}&{}C{}&0.12$\pm$0.01&0.065$\pm$0.007&0.64$\pm$0.06 \\
$^{150m}$Eu{}&{}I{}&0.51$\pm$0.05&1.51$\pm$0.15&3.93$\pm$0.4&$^{173}$Hf{}&{}C{}&--&0.05$\pm$0.007&0.17$\pm$0.02 \\
$^{151}$Nd{}&{}C{}&0.01$\pm$0.002&--&0.01$\pm$0.001&$^{175}$Hf{}&{}I{}&0.082$\pm$0.01&0.055$\pm$0.007&0.4$\pm$0.04 \\
$^{151}$Pm{}&{}I{}&0.22$\pm$0.03&0.18$\pm$0.02&0.51$\pm$0.05&$^{175}$Ta{}&{}C{}&--&--&0.012$\pm$0.002 \\
$^{153}$Sm{}&{}C{}&0.43$\pm$0.04&0.48$\pm$0.05&1.3$\pm$0.13&$^{177}$Lu{}&{}C{}&0.69$\pm$0.07&0.22$\pm$0.04&0.5$\pm$0.06 \\
$^{153}$Gd{}&{}C{}&$\leq$0.007&0.51$\pm$0.05&1.2$\pm$0.12&$^{177}$Ta{}&{}C{}&0.041$\pm$0.008&0.027$\pm$0.005&0.45$\pm$0.05 \\
$^{154}$Tb{}&{}C{}&0.04$\pm$0.005&0.22$\pm$0.04&2.1$\pm$0.21&$^{181}$Hf{}&{}C{}&0.33$\pm$0.06&0.15$\pm$0.03&0.12$\pm$0.02 \\
$^{155}$Tb{}&{}I{}&0.061$\pm$0.007&0.12$\pm$0.04&0.71$\pm$0.07&$^{181}$Re{}&{}C{}&$\leq$0.002&--&0.01$\pm$0.002 \\
$^{155}$Dy{}&{}C{}&--&0.72$\pm$0.07&0.05$\pm$0.005&$^{182}$Ta{}&{}C{}&0.56$\pm$0.06&0.23$\pm$0.02&0.5$\pm$0.05 \\
$^{156}$Tb{}&{}I{}&0.07$\pm$0.009&0.22$\pm$0.03&1.7$\pm$0.17&$^{182}$Re{}&{}C{}&$\leq$0.006&$\leq$0.005&0.03$\pm$0.006 \\
$^{157}$Eu{}&{}C{}&0.23$\pm$0.02&0.1$\pm$0.01&0.25$\pm$0.03&$^{183}$Ta{}&{}C{}&0.35$\pm$0.04&0.1$\pm$0.02&0.2$\pm$0.02 \\
$^{157}$Dy{}&{}C{}&0.012$\pm$0.002&0.016$\pm$0.002&0.11$\pm$0.02&$^{183}$Re{}&{}C{}&0.027$\pm$0.017&0.003$\pm$0.007&0.09$\pm$0.01 \\
$^{160}$Tb{}&{}I{}&0.46$\pm$0.05&0.21$\pm$0.04&1.2$\pm$0.12&$^{184}$Ta{}&{}C{}&0.21$\pm$0.03&0.041$\pm$0.008&0.08$\pm$0.01 \\
$^{167}$Ho{}&{}C{}&0.5$\pm$0.05&0.13$\pm$0.01&0.6$\pm$0.07&$^{184}$Re{}&{}I{}&0.08$\pm$0.01&0.038$\pm$0.005&0.15$\pm$0.03 \\
$^{167}$Tm{}&{}C{}&0.11$\pm$0.02&0.056$\pm$0.007&0.55$\pm$0.06&$^{188}$Ir{}&{}C{}&$\leq$0.008&$\leq$0.005&0.015$\pm$0.003 \\
$^{168}$Tm{}&{}I{}&0.15$\pm$0.03&0.1$\pm$0.02&1.52$\pm$0.15&$^{189}$Ir{}&{}C{}&0.051$\pm$0.007&0.013$\pm$0.002&0.06$\pm$0.009 \\
$^{172}$Er{}&{}C{}&0.055$\pm$0.006&0.015$\pm$0.003&0.03$\pm$0.006&$^{191}$Pt{}&{}C{}&$\leq$0.003&--&0.01$\pm$0.002 \\
\end{tabular}
\end{ruledtabular}
\end{table*}

The fragment production cross sections are usually considered as an independent yield (I) in the absence of a parent isotope (which may
give a contribution in measured cross section via $\beta^{\pm}$-decays) and are determined by using the following equation:
\begin{small} 
\begin{align}
\hspace{-0.2cm}\sigma=\frac{\Delta{N}\;\lambda}{N_{p}\,N_{n}\,k\,\epsilon\,\eta\,(1-\exp{(-\lambda
t_{1})})\exp{(-\lambda t_{2})}(1-\exp{(-\lambda t_{3})})},\label{g1}
\end{align}
\end{small}
\noindent where $\sigma$ is the cross section of the reaction fragment production (mb); $\Delta{N}$ is the area under the photopeak; $N_{p}$ is 
the projectile beam intensity (min$^{-1}$); $N_{n}$ is the number of target nuclei (in 1/cm$^{2}$ units); $t_{1}$ is the irradiation time; $t_{2}$ 
is the time of exposure between the end of the irradiation and the beginning of the measurement; $t_{3}$ is the measurement time; $\lambda$ is the 
decay constant (min$^{-1}$); $\eta$ is  the intensity of $\gamma$-transitions; $k$ is the total coefficient of $\gamma$-ray
absorption in target and detector materials, and $\epsilon$ is the $\gamma$-ray detection efficiency.

In the case where the cross section of a given isotope includes a contribution from the $\beta^{\pm}$-decay of neighboring unstable isobars, 
the cross section calculation becomes more complicated. If the formation cross section of the parent isotope is known from the experimental data, or if it 
can be estimated on the basis of other sources, the independent cross sections of daughter nuclei can be calculated by the relation \cite{Karapetyan2015}:
\begin{widetext}
\begin{eqnarray}
\sigma_{B}=&&\frac{\lambda_{B}}{(1-\exp{(-\lambda_{B}t_{1})})\,
\exp{(-\lambda_{B}t_{2})}(\,1-\exp{(-\lambda_{B} t_{3})})}\times\nonumber\\&&
\hspace*{-1.5cm}
\left.\Biggl[\frac{\Delta{N}}{N_{p}\,N_{n}\,k\,\epsilon\,\eta}-
\sigma_{A}\,f_{AB}\,\frac{\lambda_{A}\,\lambda_{B}}{\lambda_{B}-\lambda_{A}}
\Biggl(\frac{(1-\exp{(-\lambda_{A} t_{1})})\,\exp{(-\lambda_{A} t_{2})}\,(1-
\exp{(-\lambda_{A} t_{3})})}{\lambda^{2}_{A}}\right.\nonumber\\
&&\left.\qquad
-\frac{(1-\exp{(-\lambda_{B} t_{1})})\,\exp{(-\lambda_{B}
t_{2})}\,(1-\exp{(-\lambda_{B} t_{3})})}{\lambda^{2}_{B}}\Biggr)\right.\Biggr],
\label{exp_CS2}
\end{eqnarray}
\end{widetext}
\noindent where the subscripts $A$ and $B$ label variables referring to, respectively, the parent and the daughter nucleus; the coefficient $f_{AB}$ 
specifies the fraction of $A$ nuclei decaying to a $B$ nucleus ($f_{AB}=1$, when the contribution from the $\beta$-decay corresponds 100\%); and 
$\Delta{N}$ is the total photopeak area associated with the decays of the daughter and parent isotopes. The effect of the precursor can 
be negligible in some limiting cases: where the half-life of the parent nucleus is very long, or in the case where its contribution is very small. 
In the case when parent and daughter isotopes can not be separated experimentally, the calculated cross sections are classified as cumulative ones 
(C). It should be mentioned that the use of the induced-activity method imposes several restrictions on the
registration of the reaction products. For example, it is impossible to measure a stable and very short-lived, or very long-lived, isotopes. As in Equation 
(\ref{g1}), all parameters in Equantion (\ref{exp_CS2}) are determined from the experiment.

A natural uranium target of 0.164 g and 0.0487 mm thick, neptunium target of 0.742 g and 0.193 mm thick and americium target of 0.177 g and 0.043 mm thick 
were exposed to an accelerated proton beam of 660 MeV in energy from the LNR Phasotron, Joint Institute for Nuclear Research (JINR), Dubna, Russia.
The cross section for the reaction $^{27}$Al(p, 3pn)$^{24}$Na \cite{Cumming1963} at the same energy was used in monitoring the proton beam. 
The yields of fission fragments were measured in the off-line mode by the induced-activity method. The irradiation time was 27 min and the proton beam intensity 
was about $3 \times 10^{14}$ protons per min. The induced activity of the targets was measured by two detectors, an HPGe detector with efficiency of 20\% and energy resolution of 1.8 KeV (1332 KeV $^{60}$Co) and a Ge(Li) detector with efficiency of 4.8\% and energy resolution of 2.6 KeV (1332 KeV $^{60}$Co). The HPGe detector 
together with its cryostat was enclosed in a lead shield in the form of a rectangular parallelepiped of $58\times40\times29$ cm$^{3}$. The 7.5-cm-thick lead walls 
were lined with a layer of cadmium and a layer of copper, both 1 mm thick. The detection efficiencies were determined by using the standard radiation sources of 
$^{22}$Na, $^{54}$Mn, $^{57,60}$Co, and $^{137}$Cs. The spectra measurement time was gradually increased from 85 min to 14 days. Processing of the gamma-ray spectra, 
i.e. determination of the areas and positions of peaks and their detection limits against the given background, was carried out in the interactive way with the help 
of the program DEIMOS \cite{Frana2003}. Then the energy calibration refined, background lines were subtracted, and the intensity decrease periods were found 
for particular lines. Line intensities were calculated on the basis of the absolute gamma-ray detection efficiency found by measurement of calibration sources.

Further details of the experimental setup related to $^{241}$Am, $^{238}$U and $^{237}$Np reactions at 660 MeV are given in Refs. 
\cite{Deppman2013c,Karapetyan2009,Balabekyan2010}.

\section{Analysis of Fragment Mass Distributions}
\label{sec:FragDistAna}

The first step of this systematic study was to investigate each of the selected reactions at a time searching for the best possible combination of 
fission modes and best 
parameters for Equation (\ref{eqYield}). For the parameters regarding the charge distribution we adopt the best parameters already proposed in literature \cite{Karapetyan2009, Duijvestijn1999}, since the present work looks specifically at the mass distributions. 

\begin{figure}
 \centering
 \subfigure[ $^{241}$Am]{
    \includegraphics[scale=0.2,keepaspectratio=true]{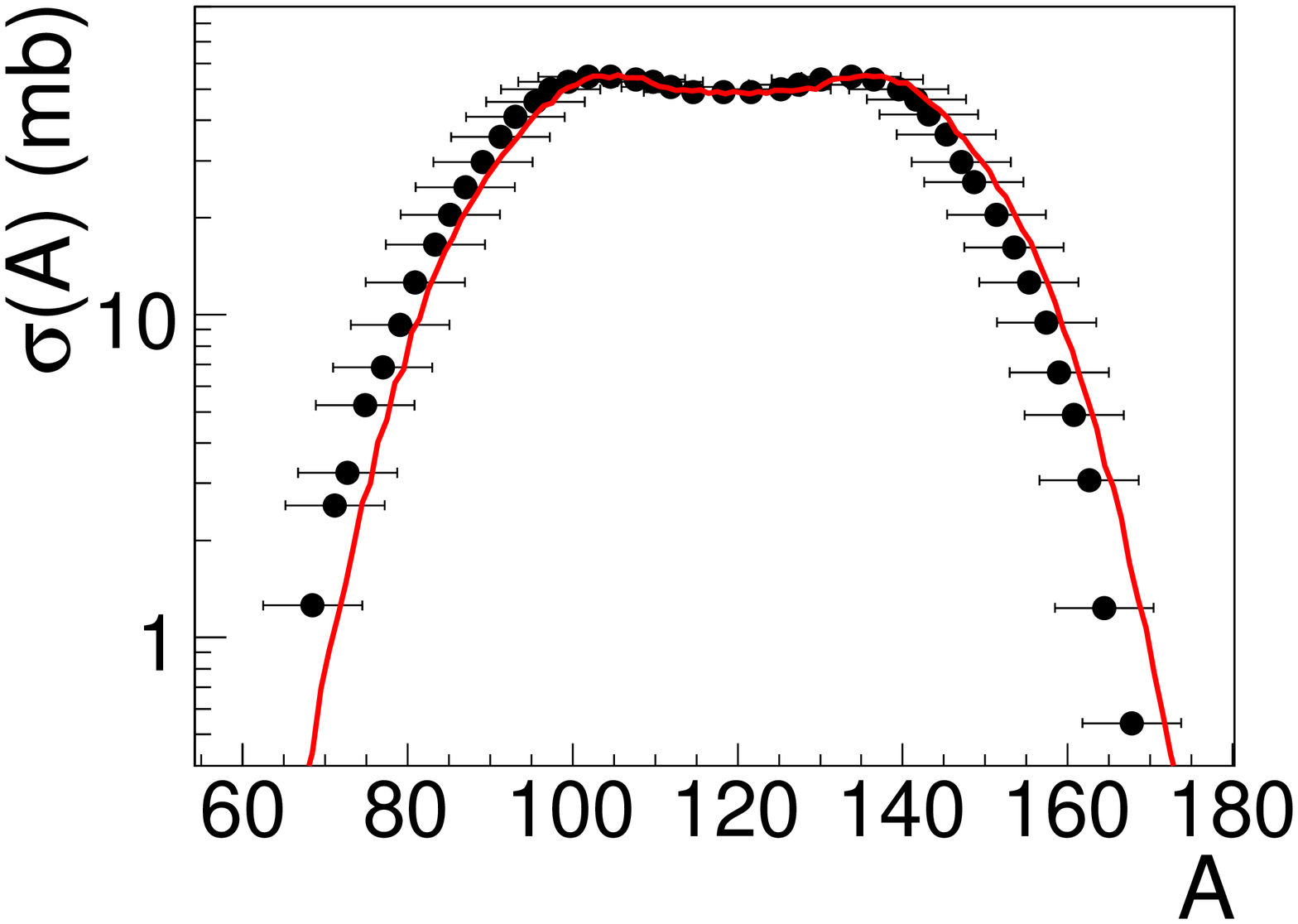}
 }
 \subfigure[ $^{237}$Np]{
    \includegraphics[scale=0.2,keepaspectratio=true]{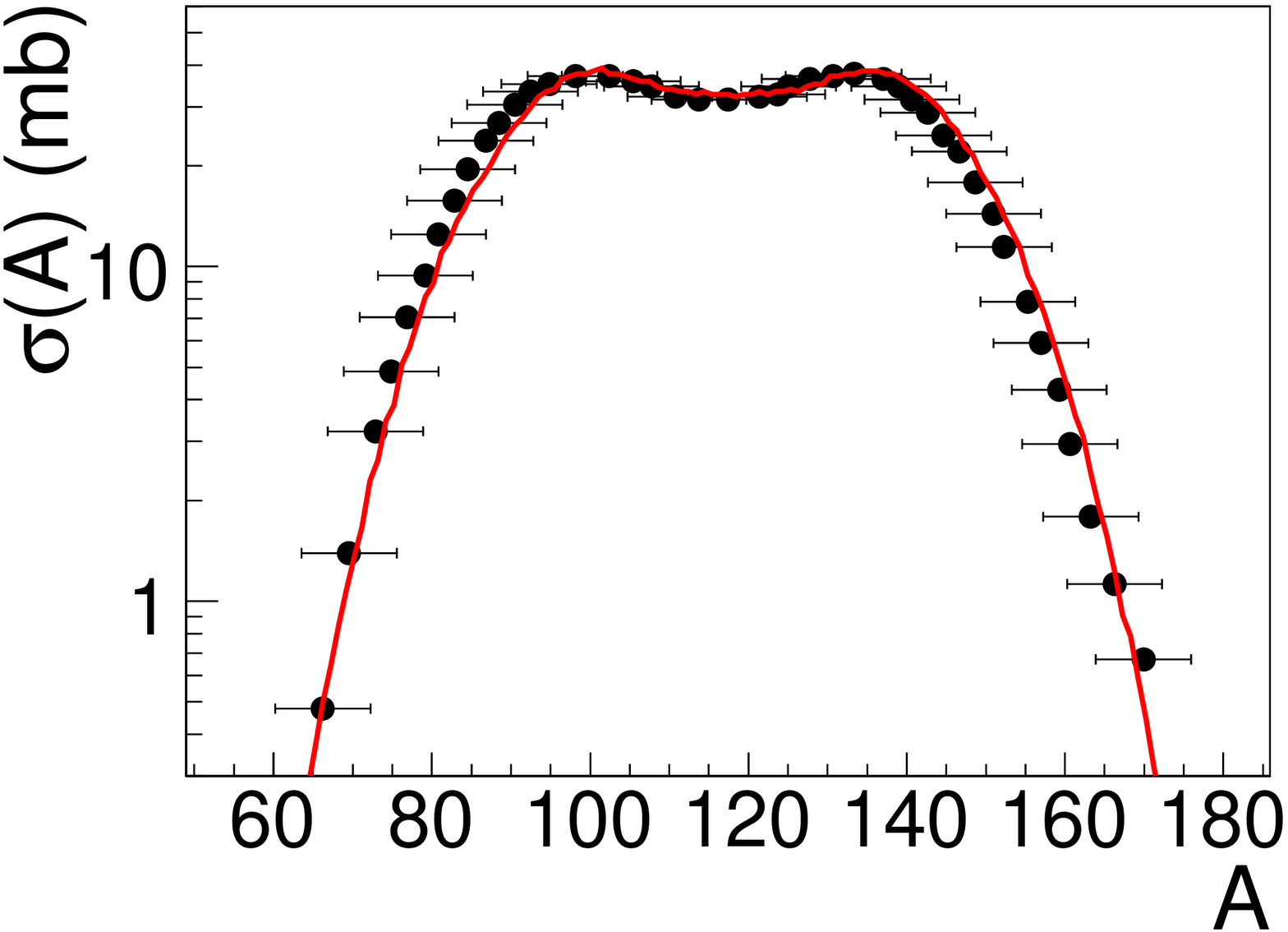}
 }
 \subfigure[ $^{238}$U]{
    \includegraphics[scale=0.2,keepaspectratio=true]{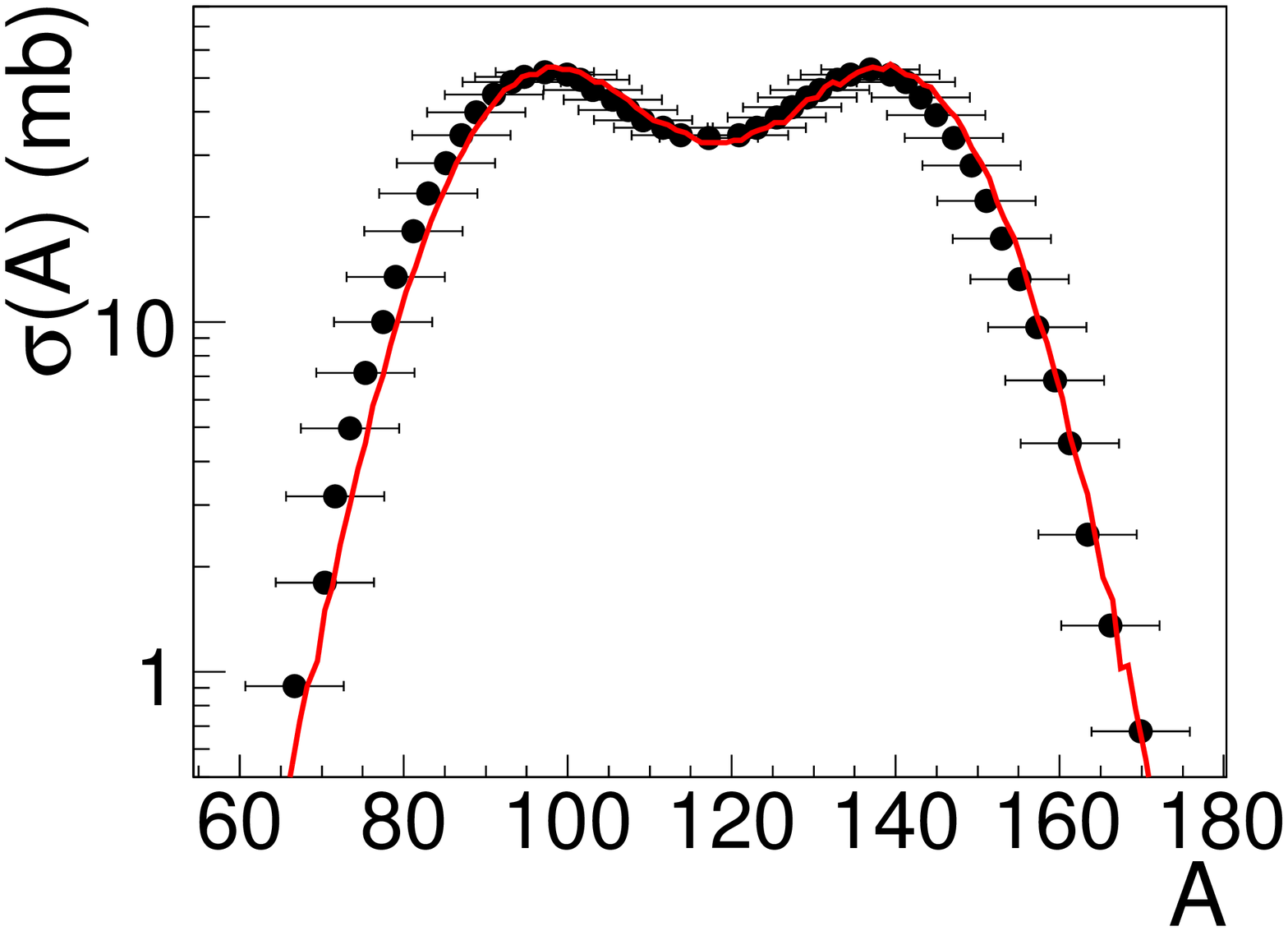}
 }
 \subfigure[ $^{232}$Th]{
    \includegraphics[scale=0.2,keepaspectratio=true]{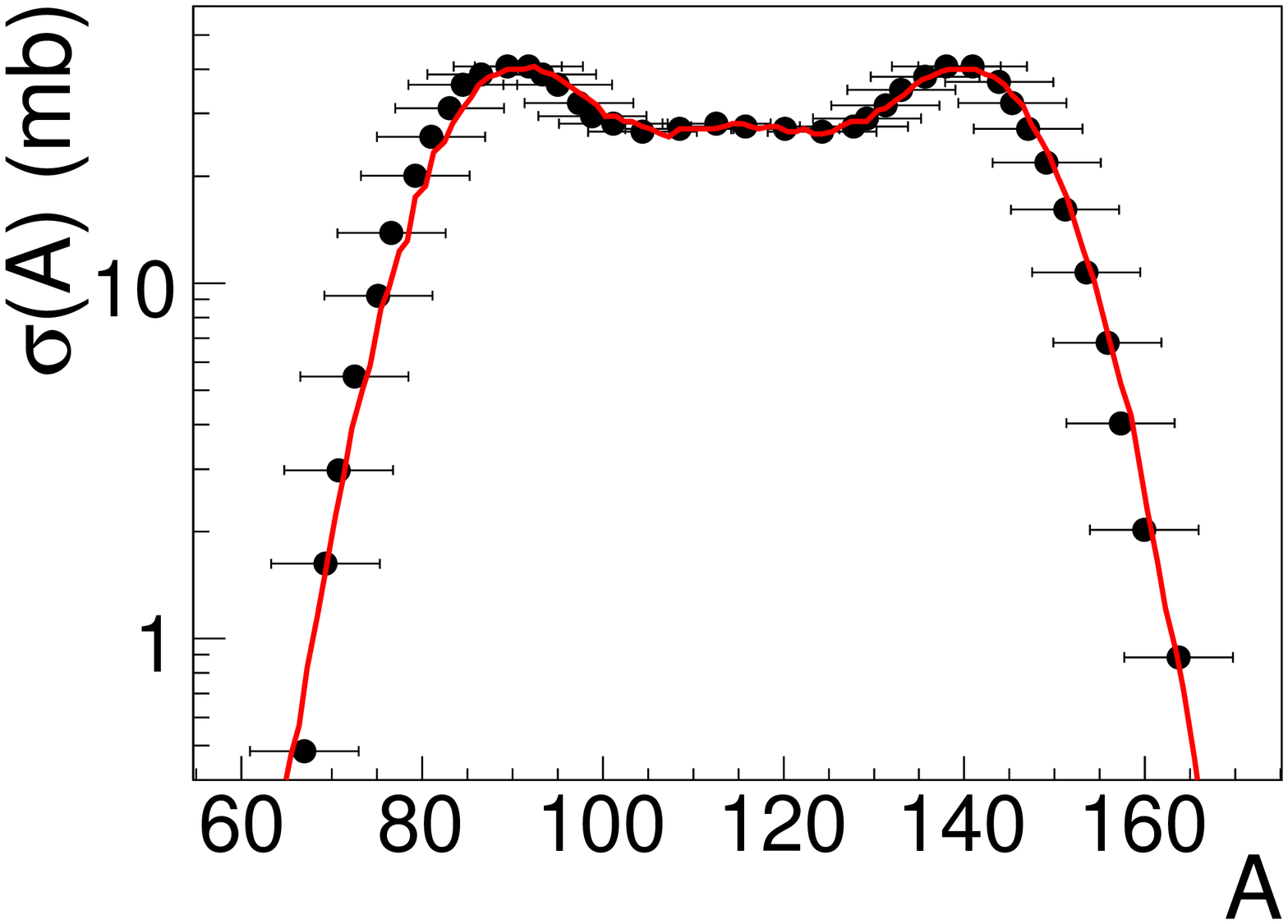}
 }
\caption{(Color online) Fragment mass distributions from the fission of a) $^{241}$Am b) $^{237}$Np c) $^{238}$U and d) $^{232}$Th by a 26.5 MeV proton. 
Solid line represents CRISP calculation. Experimental data from \cite{Demetriou2010}.}
\label{fig26MeV}
\end{figure}

\begin{figure}
 \centering
 \subfigure[ $^{241}$Am]{
    \includegraphics[scale=0.2,keepaspectratio=true]{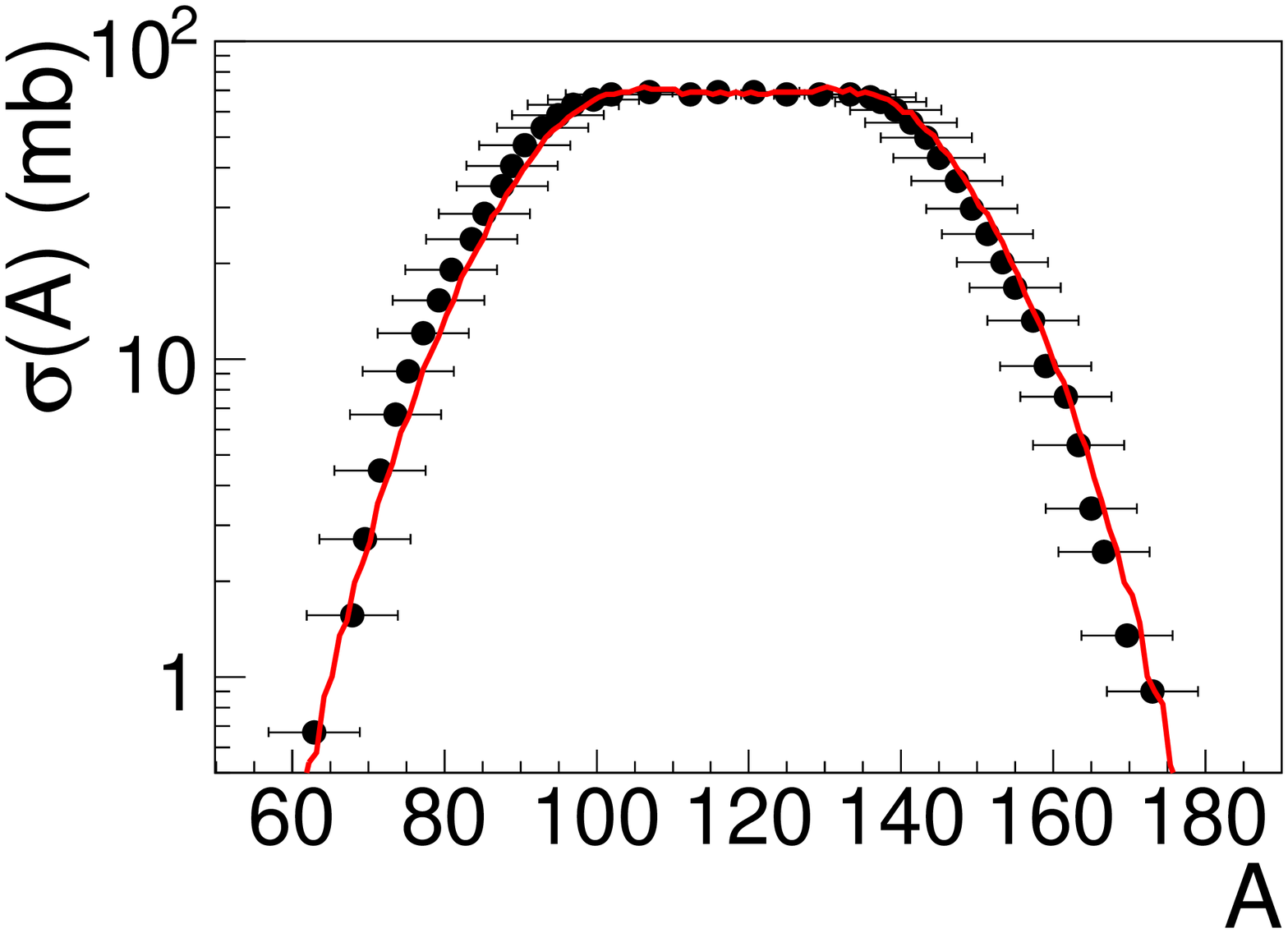}
 }
 \subfigure[ $^{237}$Np]{
    \includegraphics[scale=0.2,keepaspectratio=true]{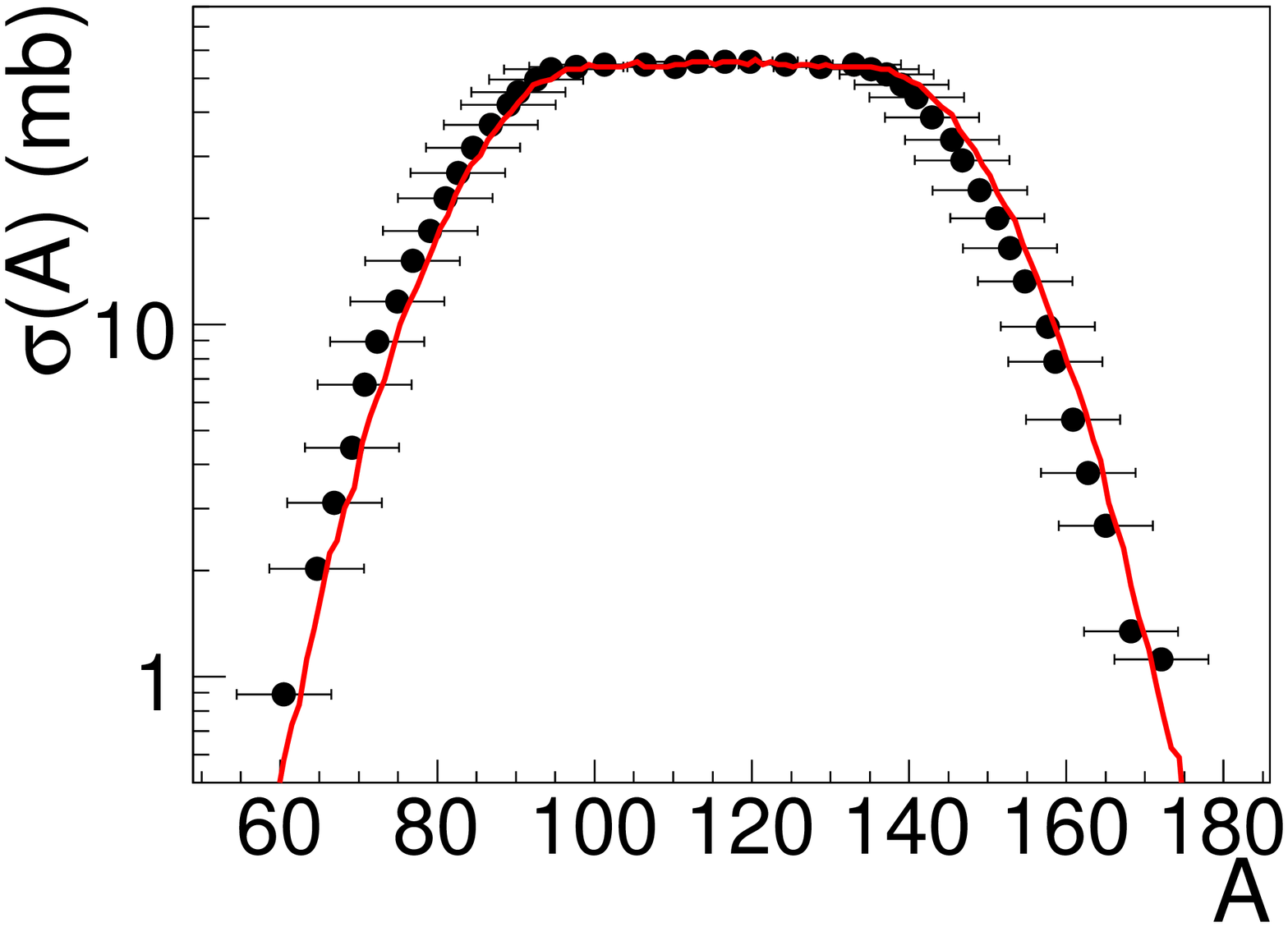}
 }
 \subfigure[ $^{238}$U]{
    \includegraphics[scale=0.2,keepaspectratio=true]{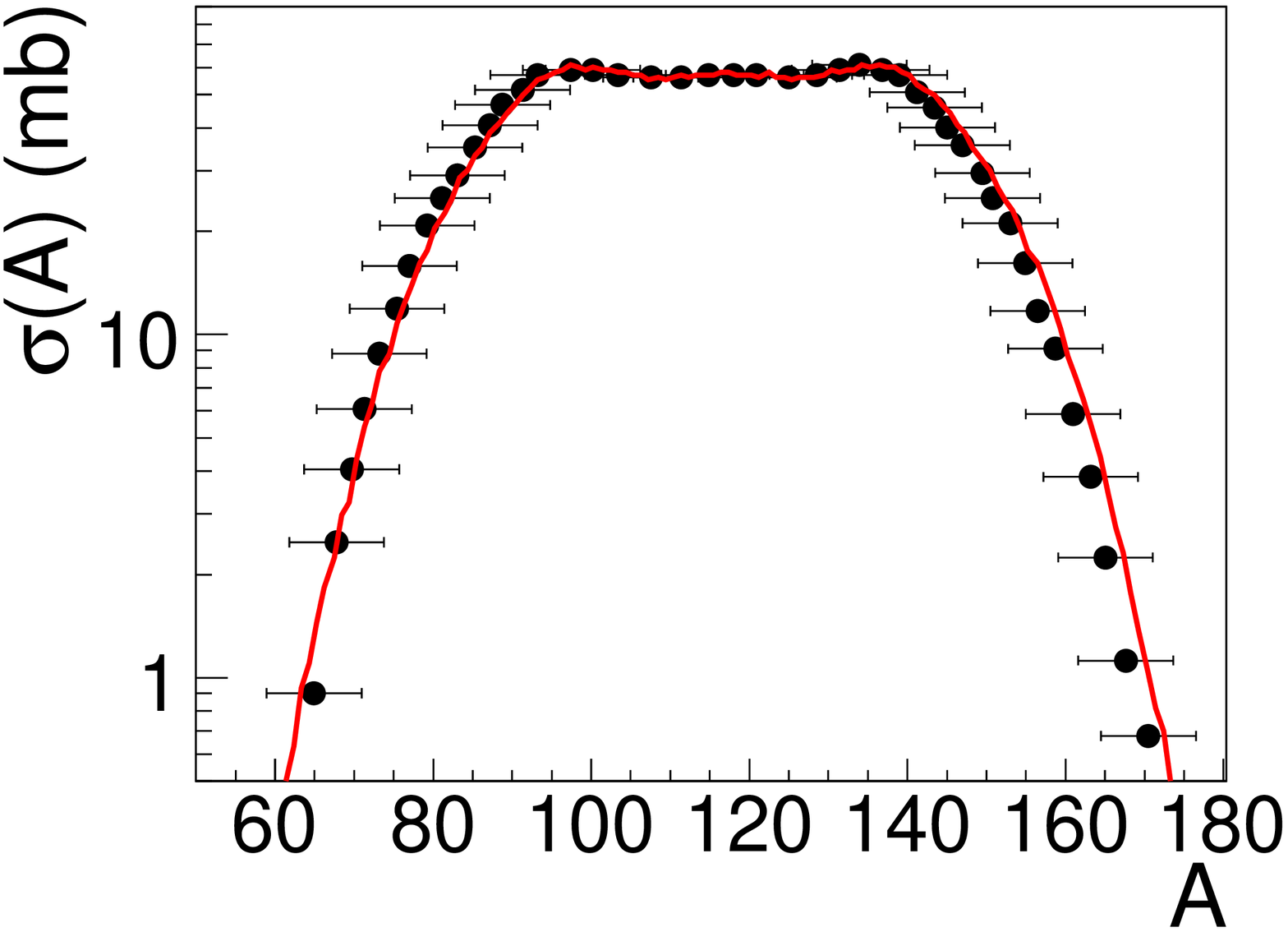}
 }
 \subfigure[ $^{232}$Th]{
    \includegraphics[scale=0.2,keepaspectratio=true]{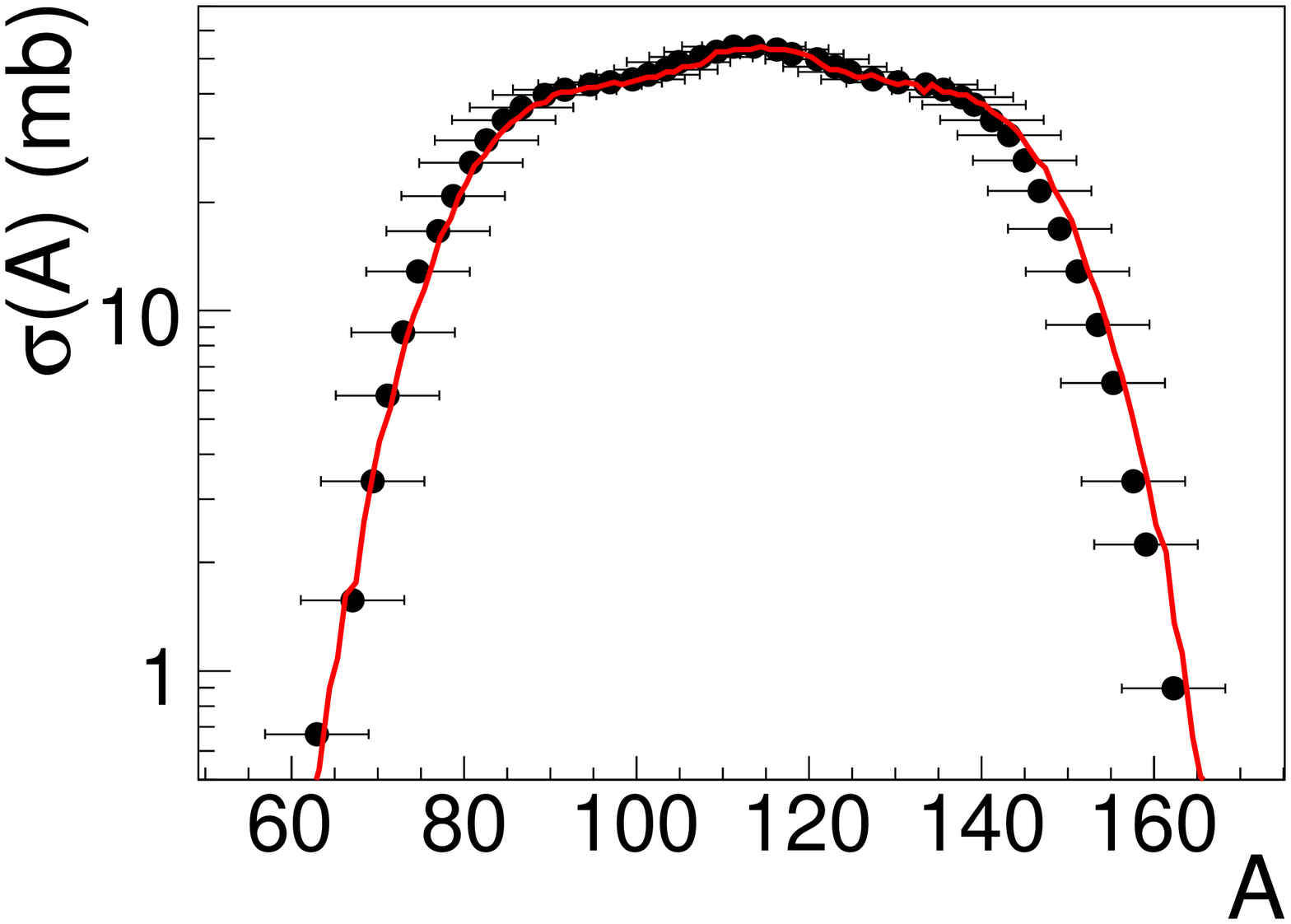}
 }
\caption{(Color online) Fragment mass distributions from the fission of a) $^{241}$Am b) $^{237}$Np c) $^{238}$U and d) $^{232}$Th by a 62.9 MeV proton. 
Solid line represents CRISP calculation. Experimental data from \cite{Demetriou2010}.}
\label{fig62MeV}
\end{figure}

\begin{figure}
 \centering
 \subfigure[ $^{241}$Am]{
    \includegraphics[scale=0.2,keepaspectratio=true]{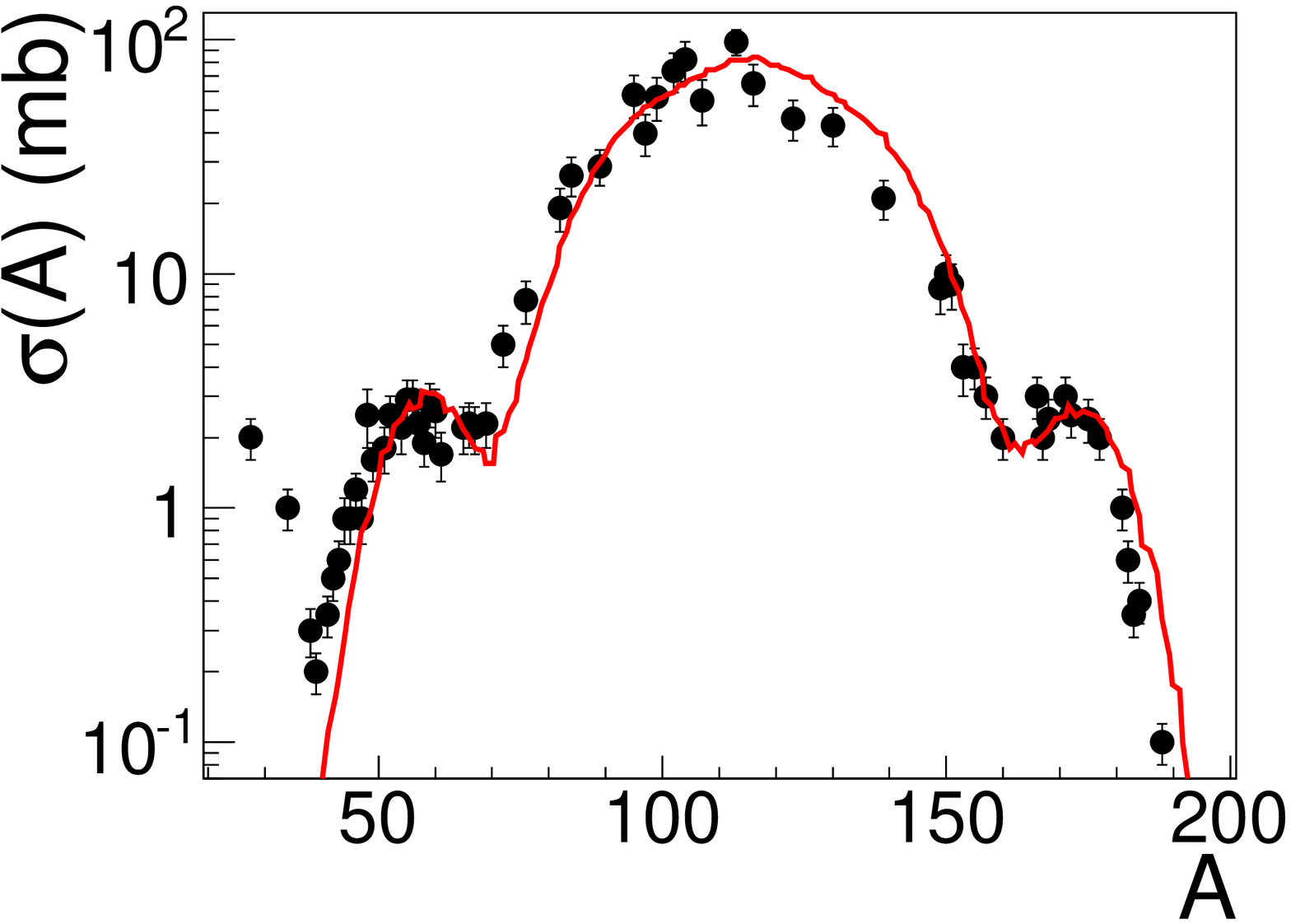}
    \label{figAm660}
 }
 \subfigure[ $^{237}$Np]{
    \includegraphics[scale=0.2,keepaspectratio=true]{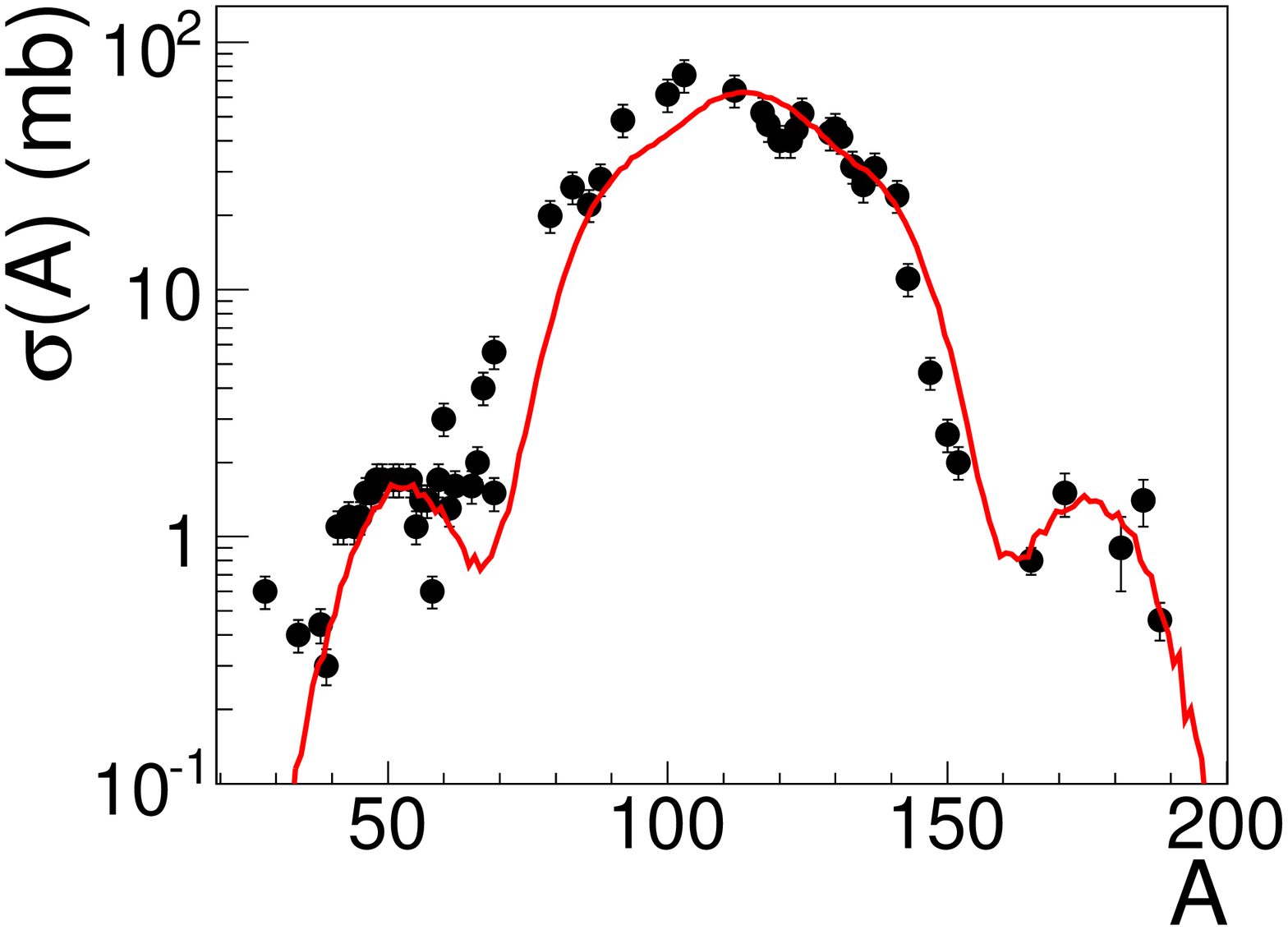}
    \label{figNp660}
 }
 \subfigure[ $^{238}$U]{
    \includegraphics[scale=0.2,keepaspectratio=true]{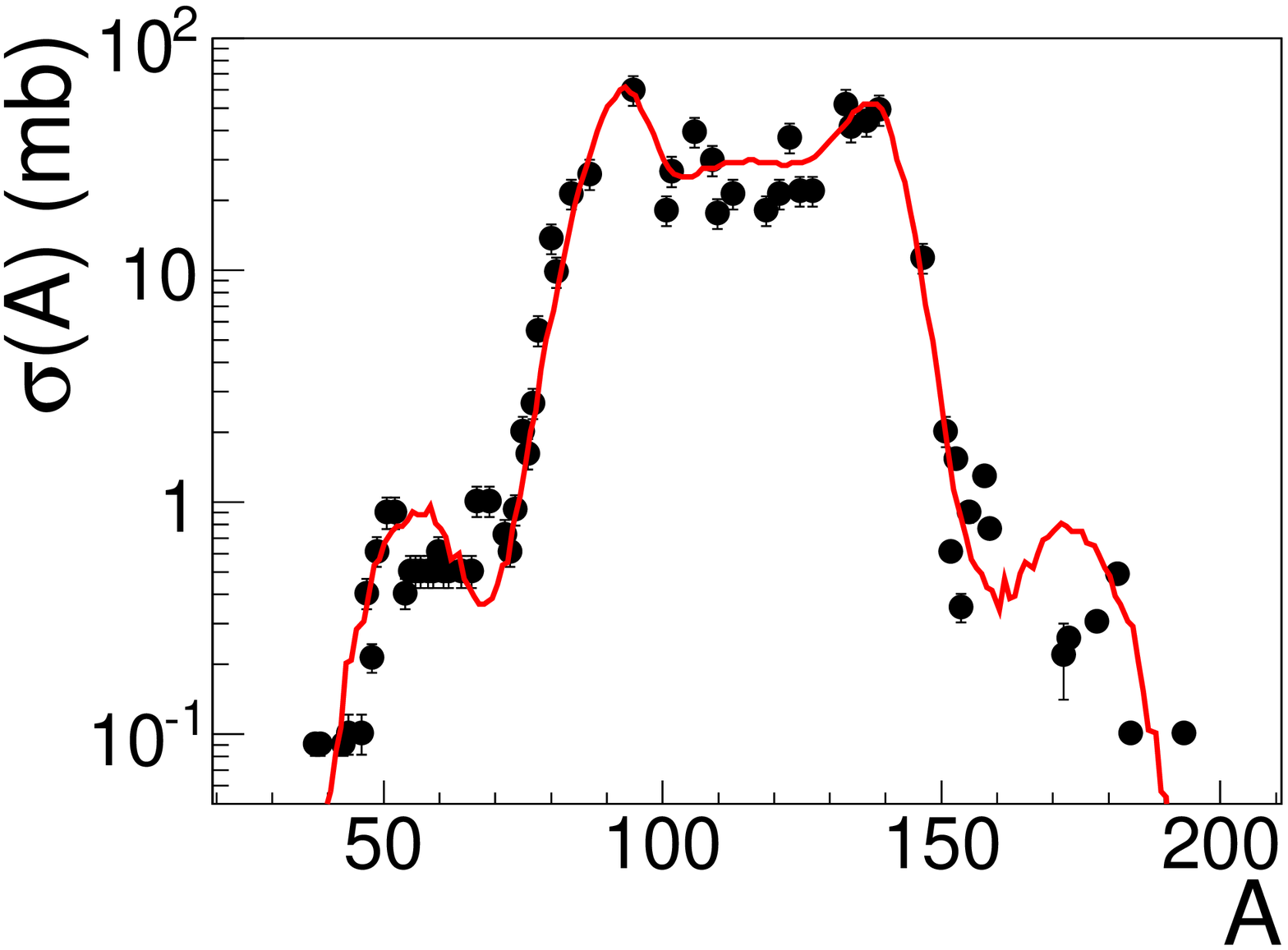}
    \label{figU660}
 }
 \subfigure[ $^{232}$Th]{
    \includegraphics[scale=0.2,keepaspectratio=true]{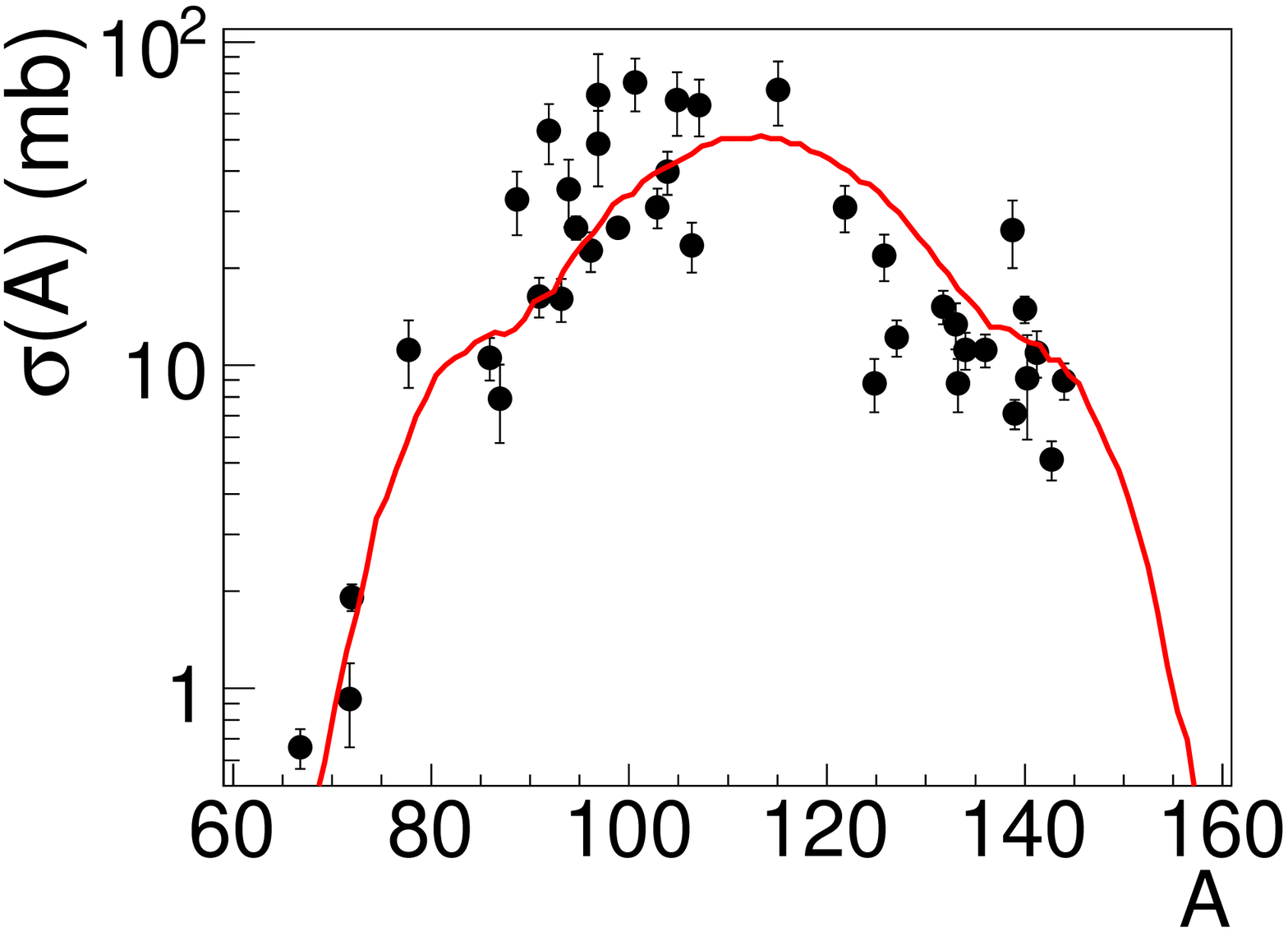}
    \label{fig660_190MeV:Th232}
 }
\caption{(Color online) Fragment mass distributions from the fission of a) $^{241}$Am b) $^{237}$Np and c) $^{238}$U by a 660 MeV proton and of 
d) $^{232}$Th by a 190 MeV proton. Solid line represents CRISP calculation. Experimental data from \cite{Karapetyan2009, Balabekyan2010, Duijvestijn1999}.}
\label{fig660_190MeV}
\end{figure}

\begin{figure}
 \centering
 \subfigure[ $^{208}$Pb + 190 MeV p]{
    \includegraphics[scale=0.2,keepaspectratio=true]{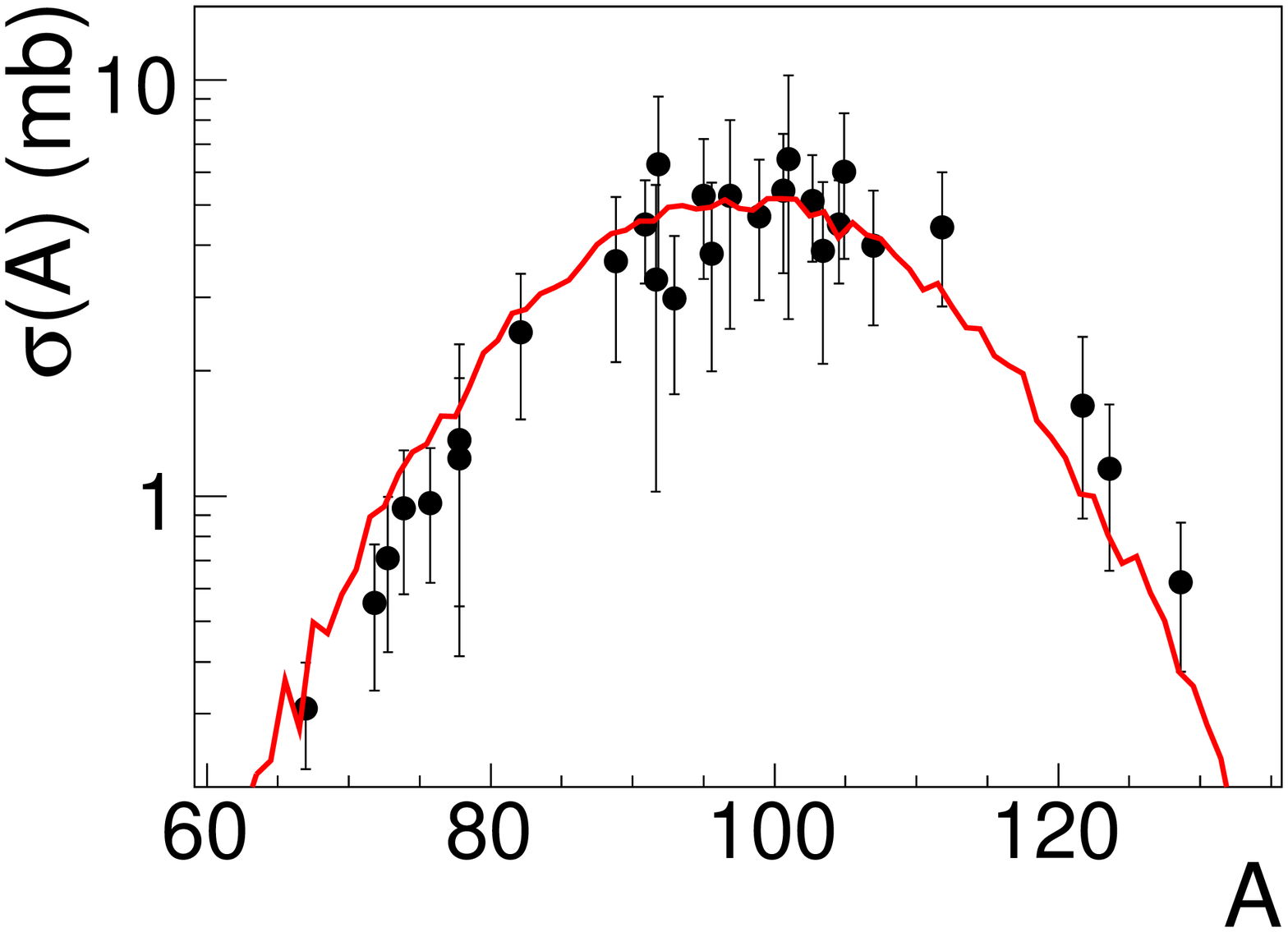}
    \label{figPbAu:Pb190}
 }
 \subfigure[ $^{208}$Pb + 500 MeV p]{
    \includegraphics[scale=0.2,keepaspectratio=true]{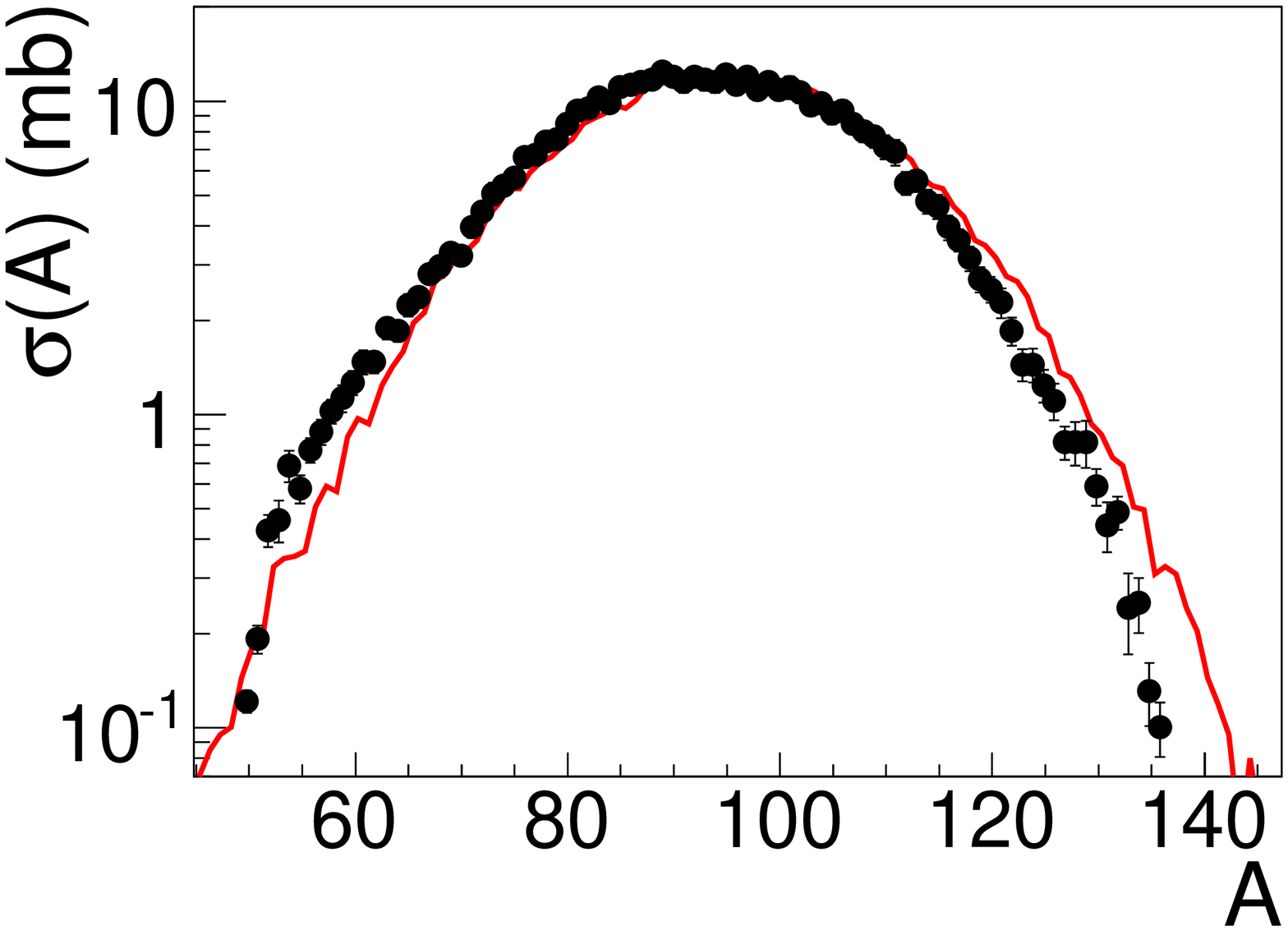}
 }
 \subfigure[ $^{208}$Pb + 1000 MeV p]{
    \includegraphics[scale=0.2,keepaspectratio=true]{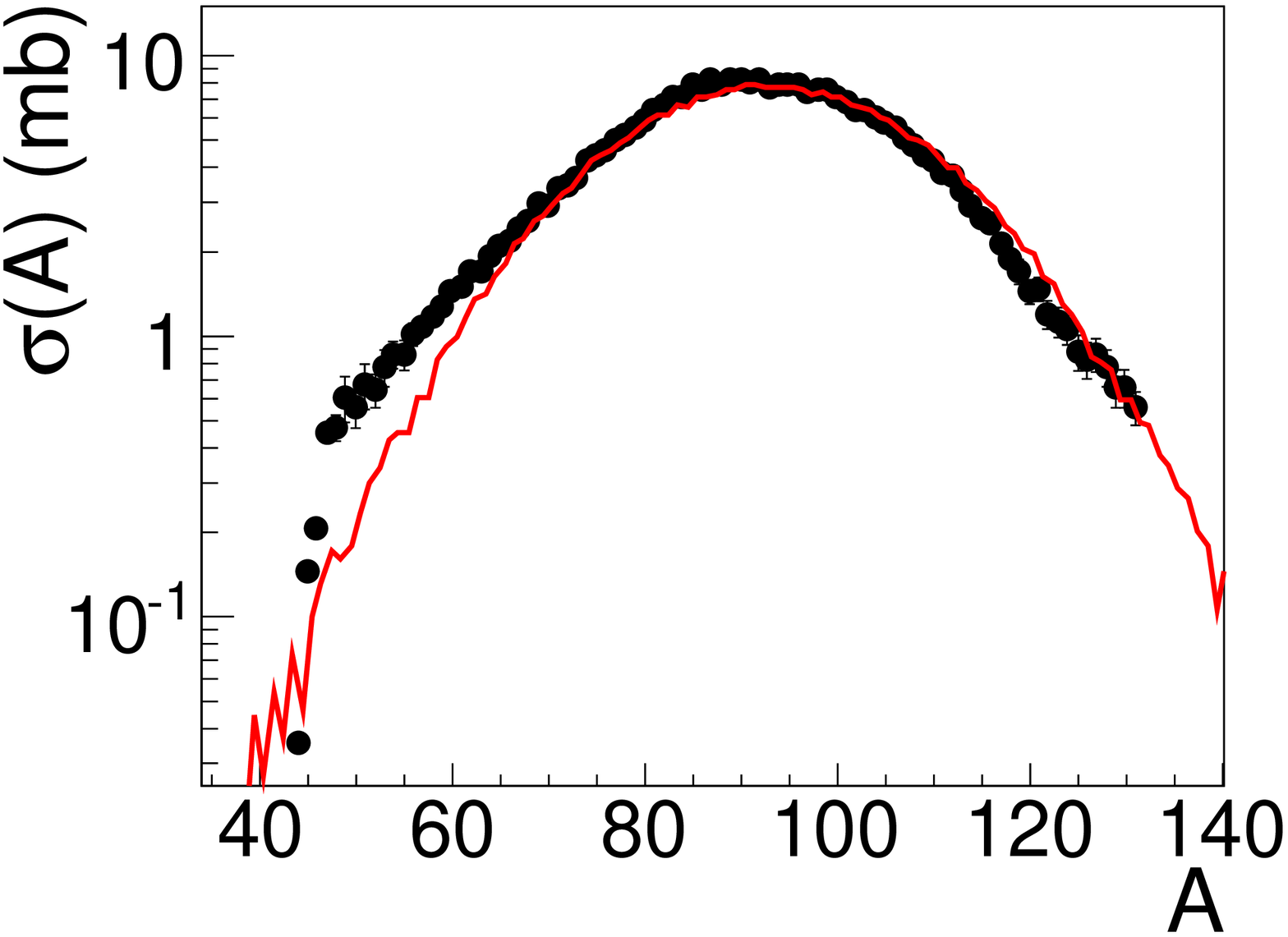}
    \label{figPbAu:Pb1000}
 }
 \subfigure[ $^{197}$Au + 190 MeV p]{
    \includegraphics[scale=0.2,keepaspectratio=true]{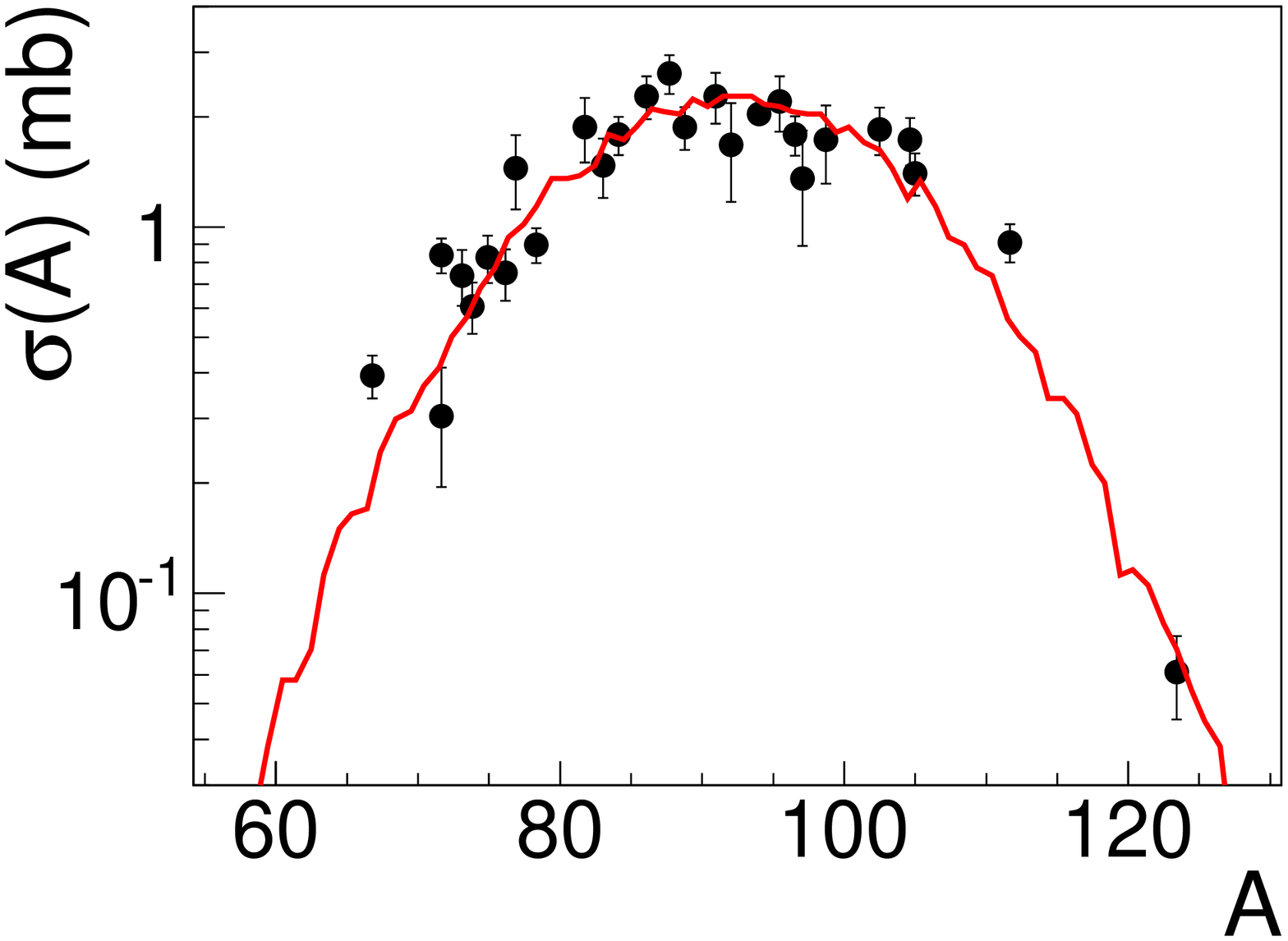}
 }
 \subfigure[ $^{197}$Au + 800 MeV p]{
    \includegraphics[scale=0.2,keepaspectratio=true]{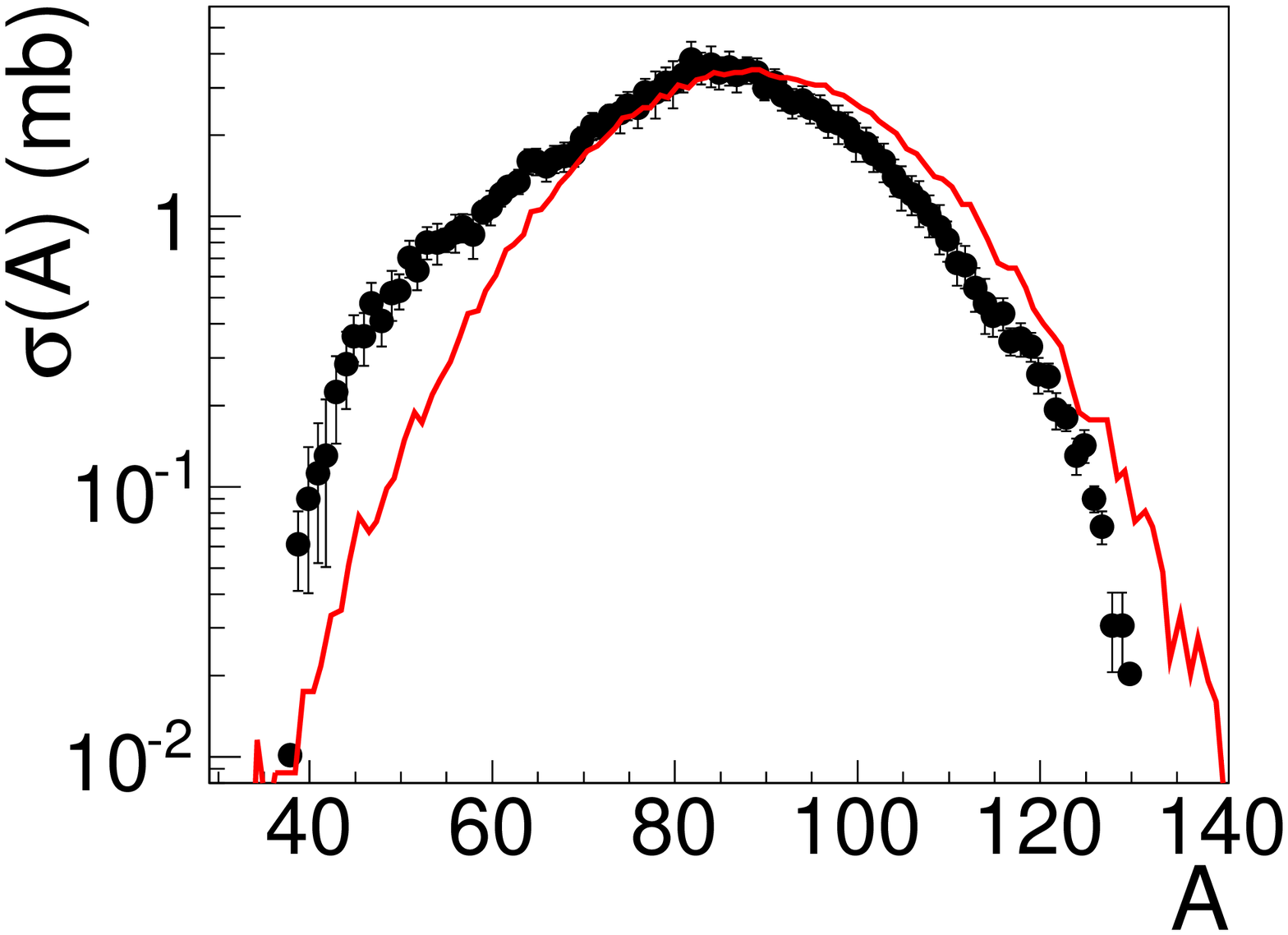}
    \label{figPbAu:Au800}
 }
\caption{(Color online) Fragment mass distributions from fission reactions a) $^{208}$Pb + 190 MeV p b) $^{208}$Pb + 500 MeV p c) $^{208}$Pb + 1000 MeV p 
 d) $^{197}$Au + 190 MeV p and e) $^{197}$Au + 800 MeV p. Solid line represents CRISP calculation. Experimental data from \cite{Duijvestijn1999,Enqvist2001,Dominguez2005,Benlliure2001}.}
\label{figPbAu}
\end{figure}

Figures \ref{fig26MeV}-\ref{figPbAu} show mass distributions of fission fragments from reactions with different target nuclei and different 
energies of the proton in comparison with experimental distributions. 

Data from Figures \ref{fig26MeV} and \ref{fig62MeV} were 
published by Demetriou et al \cite{Demetriou2010}. Data at 660 MeV from Figure \ref{fig660_190MeV} were obtained as explained above in Section \ref{sec:ExpRes}.
Those in Figure \ref{fig660_190MeV:Th232} were measured by Duijvestijn et al \cite{Duijvestijn1999}.
By Duijvestijn et al also are the experimental data of Figure \ref{figPbAu} for 190 MeV proton. Reactions $^{208}$Pb + 500 MeV p and $^{208}$Pb + 1 GeV p were published 
by Domingu\'ez et al \cite{Dominguez2005} and Enqvist et al \cite{Enqvist2001}, respectively. Benlliure et al \cite{Benlliure2001} measured the 
reaction $^{197}$Au + 800 MeV p. 

A specific set of values for the multi-modal fission parameters was determined for each reaction.

A general observation shows that to a higher or a lower degree, all calculations from Figures \ref{fig26MeV}-\ref{figPbAu} are 
shifted to the right with respect to the data, just one or two mass units in most of the cases. This systematic deviation only means that the evaporation 
chain is ending one or two steps earlier which is actually remarkable considering that the same statistical evaporation-fission competition model is used 
for all reactions. The only exception is the reaction $^{208}$Pb + 190 MeV p (Figure \ref{figPbAu:Pb190}) where the evaporation rate was higher.

The shape of all distributions are very well described except for Figures \ref{figPbAu:Pb1000} and \ref{figPbAu:Au800}. For the latter, the experimental distribution is clearly not symmetric. 
Even accounting for fragment evaporation, CRISP model cannot describe the cross section of lower masses for the $^{197}$Au + 800 MeV p reaction.

Regarding the fragments with mass number $A\lesssim 70$ shown in Figures \ref{figAm660}, \ref{figNp660} and \ref{figU660} there is an assumption that one possible 
source of production in that mass range is the spallation process with the emission of IMFs, the so-called associated spallation \cite{Yariv1979}.  

As a matter of fact, this structure in the mass-yield distribution can be understood as a result of the 
influence of shell closure at $Z=28$, which is the only magic proton shell to be expected in the light-fragment mass distribution, 
corroborating the hypothesis in Refs. \cite{Chung1981,Chung1982} of structure affecting fission even at high energies. In a previous study of $^{238}$U 
and $^{237}$Np proton-induced fission at 660 MeV \cite{Deppman2013c} by some of the authors, the production mechanism for IMFs in the mass range of $20 < A < 70$ 
was discussed in the frame of the super-asymmetric fission mechanism. The observation of another shoulder in region $170 < A < 200$ reinforces 
the idea of a fission process as the origin of those IMFs.

\begin{figure}
 \centering
 \subfigure[50 MeV]{
    \includegraphics[scale=0.2,keepaspectratio=true]{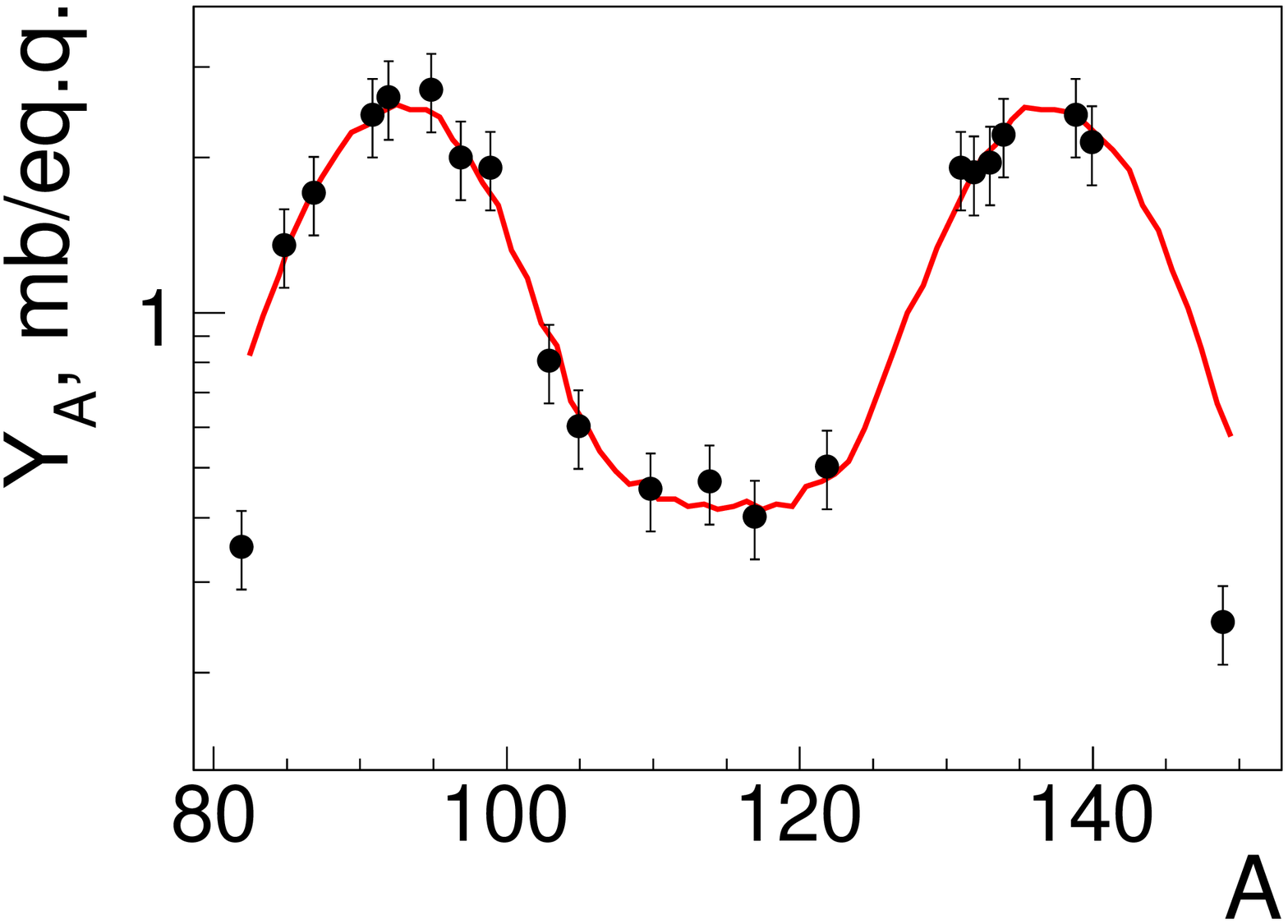}
 }
 \subfigure[3500 MeV]{
    \includegraphics[scale=0.2,keepaspectratio=true]{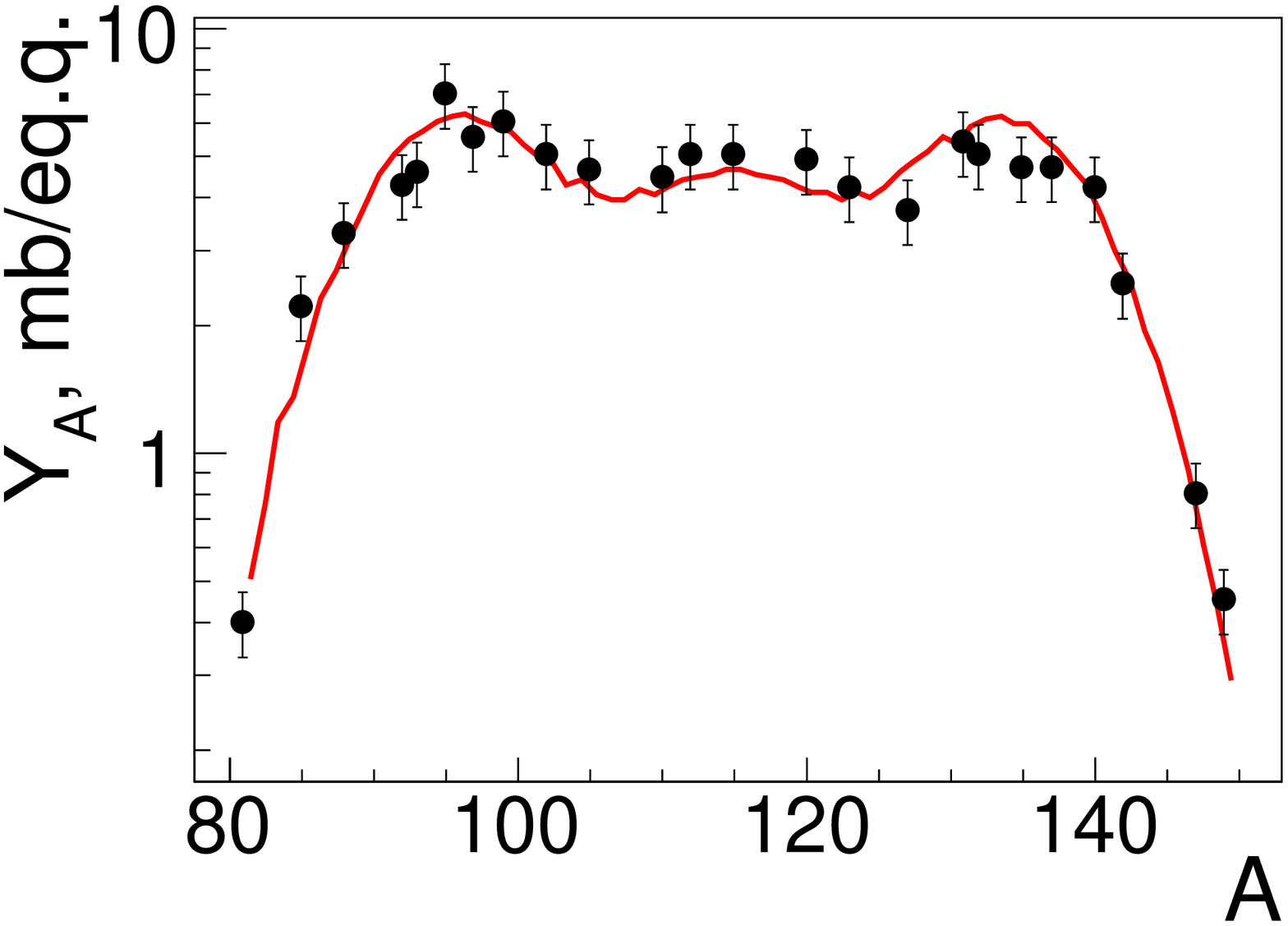}
 }
 \caption{(Color online) Fission fragment mass distributions from the Bremsstrahlung reaction of $^{232}$Th at a) 50 MeV and b) 3500 MeV maximum energies. 
Solid line represents CRISP calculation. Experimental data from \cite{Demekhina2010}.}
 \label{figBremsTh}
\end{figure}

\begin{figure}
 \centering
 \subfigure[50 MeV]{
    \includegraphics[scale=0.2,keepaspectratio=true]{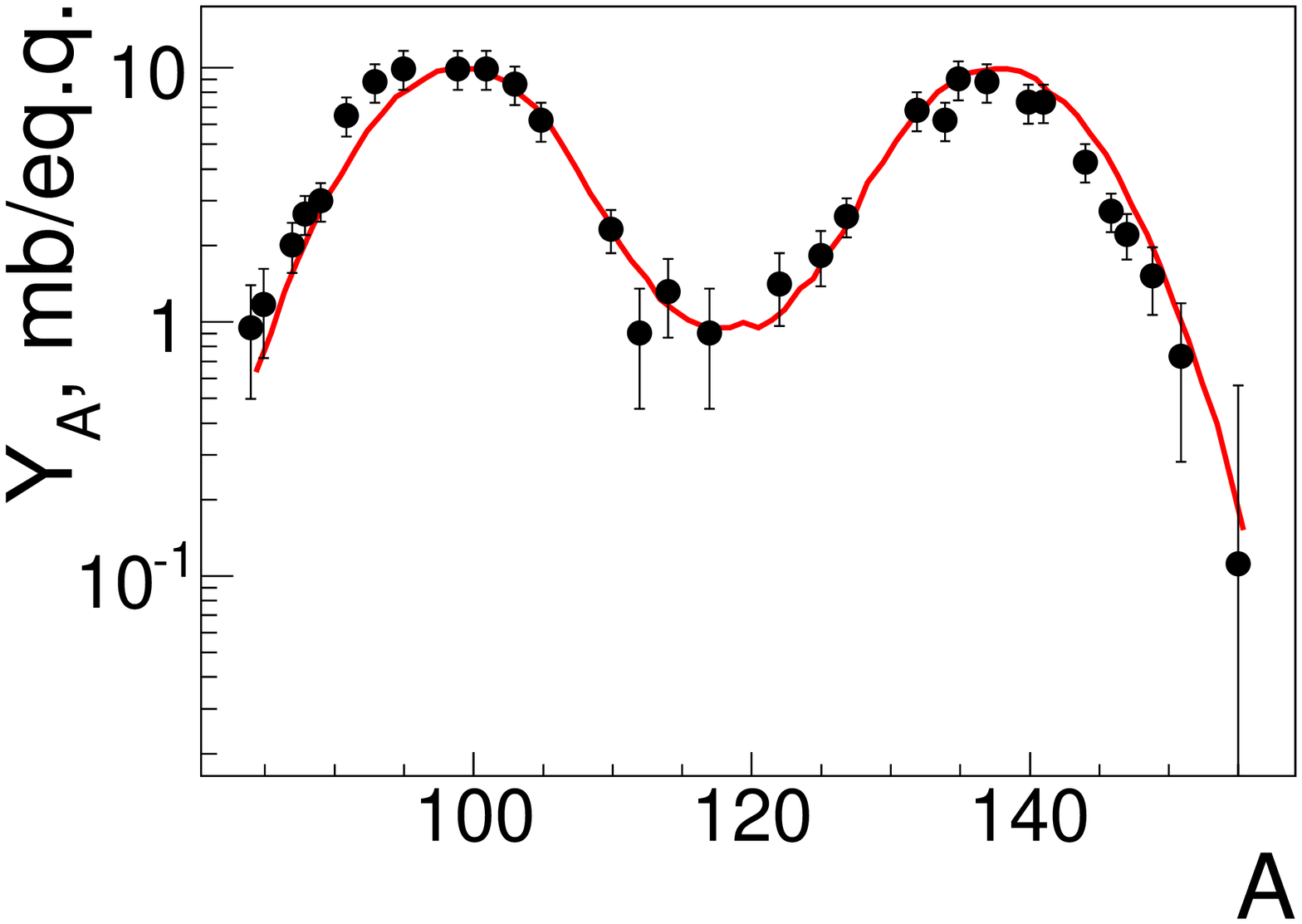}
 }
 \subfigure[3500 MeV]{
    \includegraphics[scale=0.2,keepaspectratio=true]{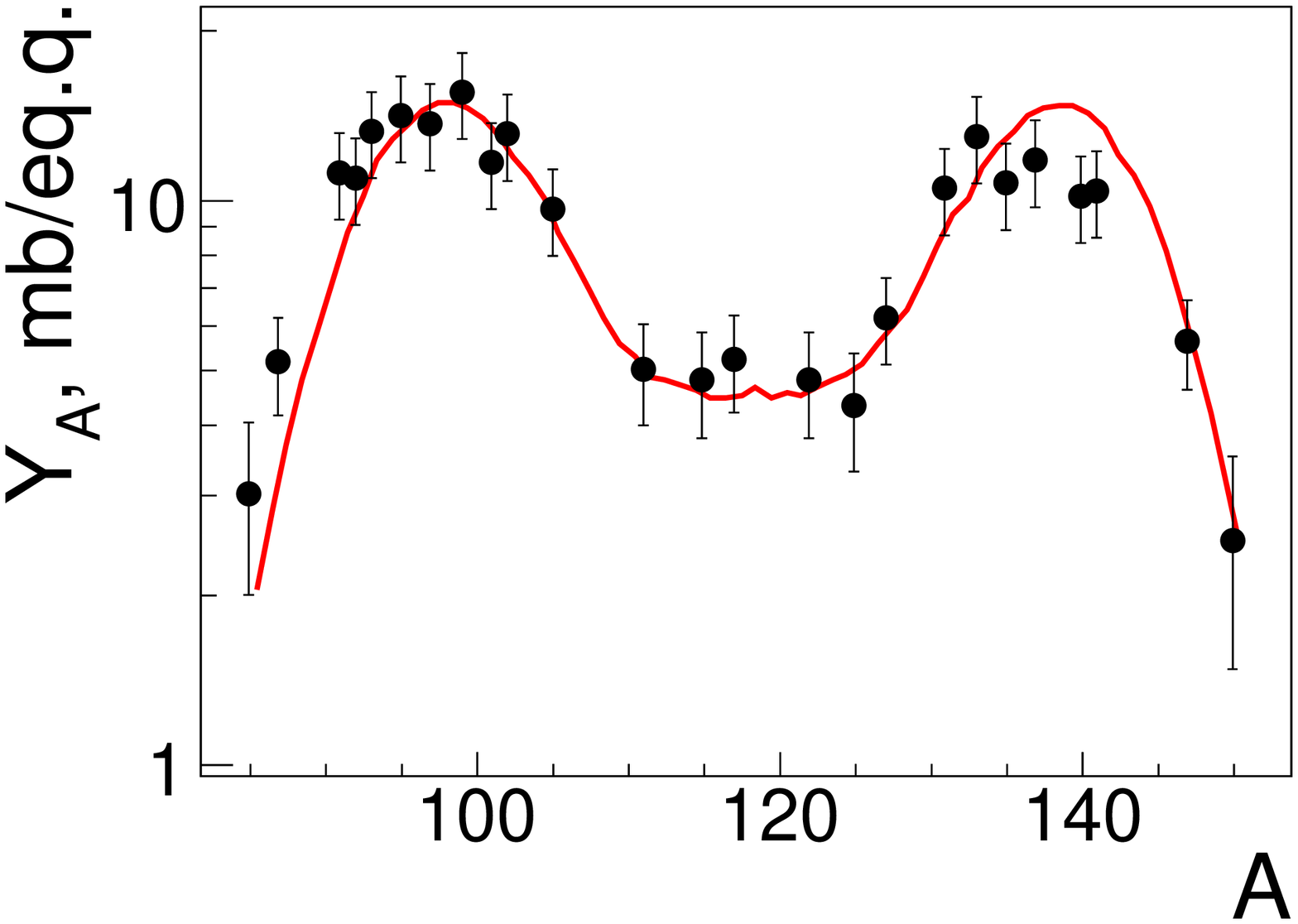}
 }
 \caption{(Color online) Fission fragment mass distributions from the Bremsstrahlung reaction of $^{238}$U at a) 50 MeV and b) 3500 MeV maximum energies. 
Solid line represents CRISP calculation. Experimental data from \cite{Demekhina2008}.}
 \label{figBremsU}
\end{figure}

Fragment mass distributions from the Bremsstrahlung reaction of $^{232}$Th and $^{238}$U, respectively, at the 
maximum energies of 50 MeV and 3500 MeV in comparison with the experimental data from \cite{Demekhina2010, Demekhina2008} are shown in Figures \ref{figBremsTh} and \ref{figBremsU}.

The deviation of the calculated distribution to the right is present only for the $^{238}$U target for both maximum energies and to the same degree as for the previous 
reactions, all initiated by a proton. Just as before, the shape of all distributions agree very closely with the experimental points.

\section{Systematic Analysis of the Fission Modes} \label{sec:SysAnaFissMod}

What follows is an analysis of how the parameters of the fission modes behave according to the excitation energy of the fissioning system. The normalization 
parameters ($K's$) don't refer to cross section. Instead, the meaningful quantity is the probability

\begin{align}
 P_i = \dfrac{K_i}{K_{\rm SL} + 2K_{\rm S1} + 2K_{\rm S2} + 2K_{\rm S3}},
\end{align}

\noindent where the index $i$ now denotes all fission modes and the factor 2 accounts for the heavy and light fragments contribution.

\begin{figure}
 \centering
 \subfigure[Superlong]{
    \includegraphics[scale=0.34,keepaspectratio=true]{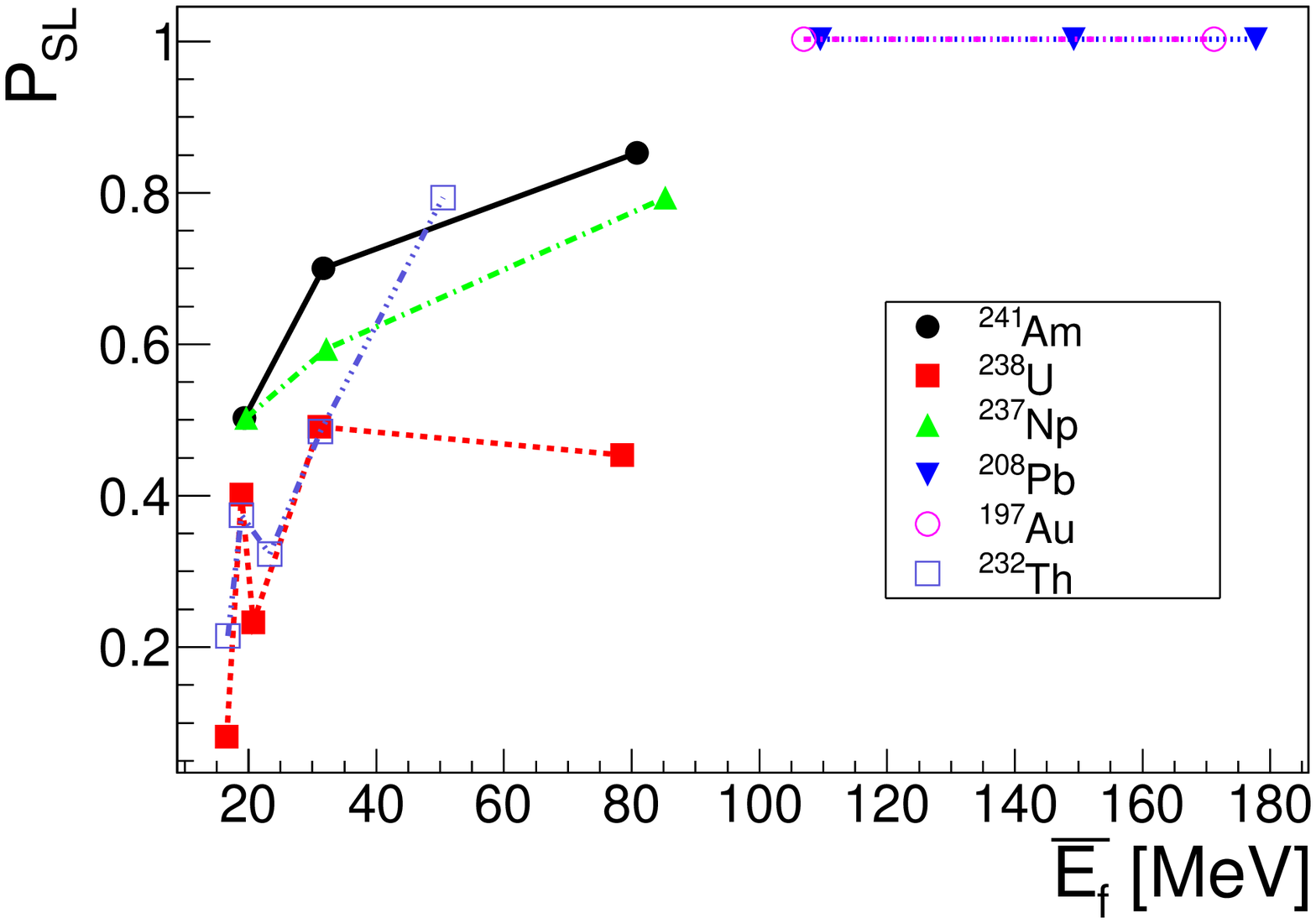}
    \label{figBehaK:SL}
 }
 \subfigure[Standard 1]{
    \includegraphics[scale=0.34,keepaspectratio=true]{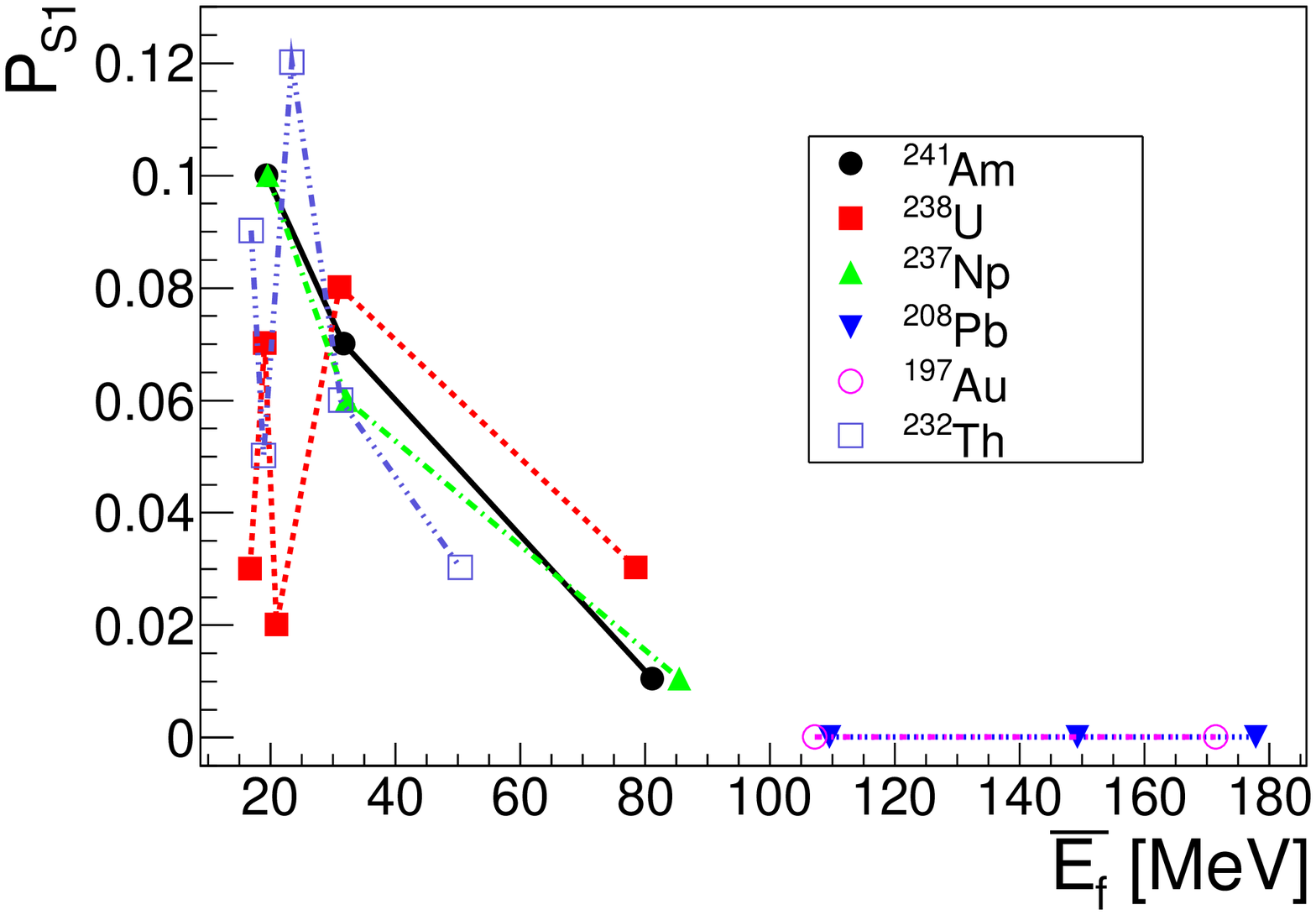}
    \label{figBehaK:S1}
 }
 \subfigure[Standard 2]{
    \includegraphics[scale=0.34,keepaspectratio=true]{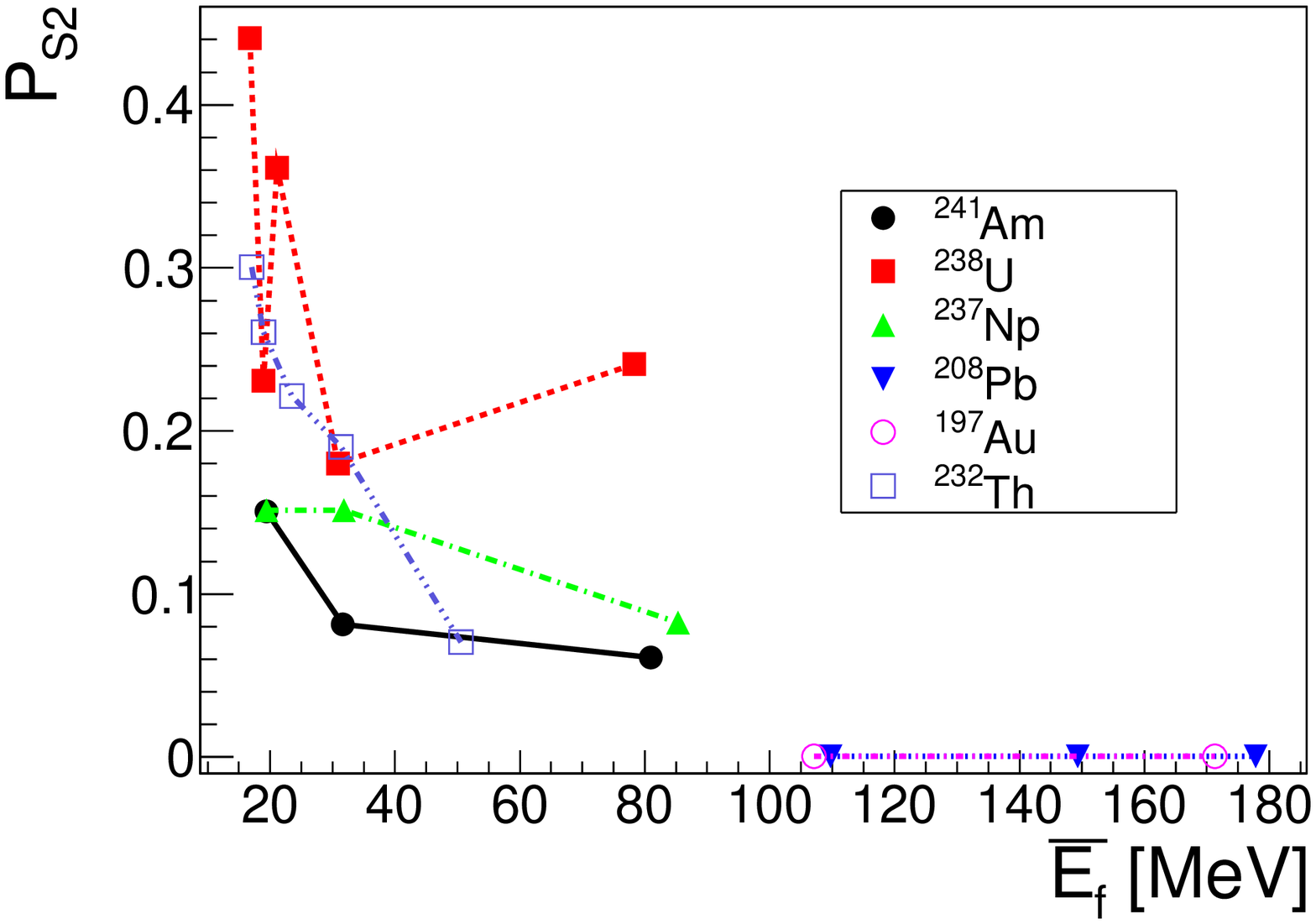}
    \label{figBehaK:S2}
 }
 \caption{(Color online) Behavior of the probability $P_i$ for the three fission modes according to the average excitation energy $\overline{E_f}$ of the 
fissioning system. a) Superlong b) Standard 1 and c) Standard 2. Fission modes probabilities have a 15\% uncertainty.}
 \label{figBehaK}
\end{figure}

The behavior of $P_i$ for the fission modes applied to reproduce all previously presented mass distributions according 
to the average excitation energy of the fissioning system is shown in Figure \ref{figBehaK}. One can see that, in general, the probability of 
symmetric fission (Fig. \ref{figBehaK:SL}) increases with increasing excitation energy of the fissioning system while the probability of asymmetric fission decreases, showing once 
more the well-known correlation between excitation energy and symmetric fission. 

\begin{figure}
 \centering
 \subfigure[Superlong]{
    \includegraphics[scale=0.34,keepaspectratio=true]{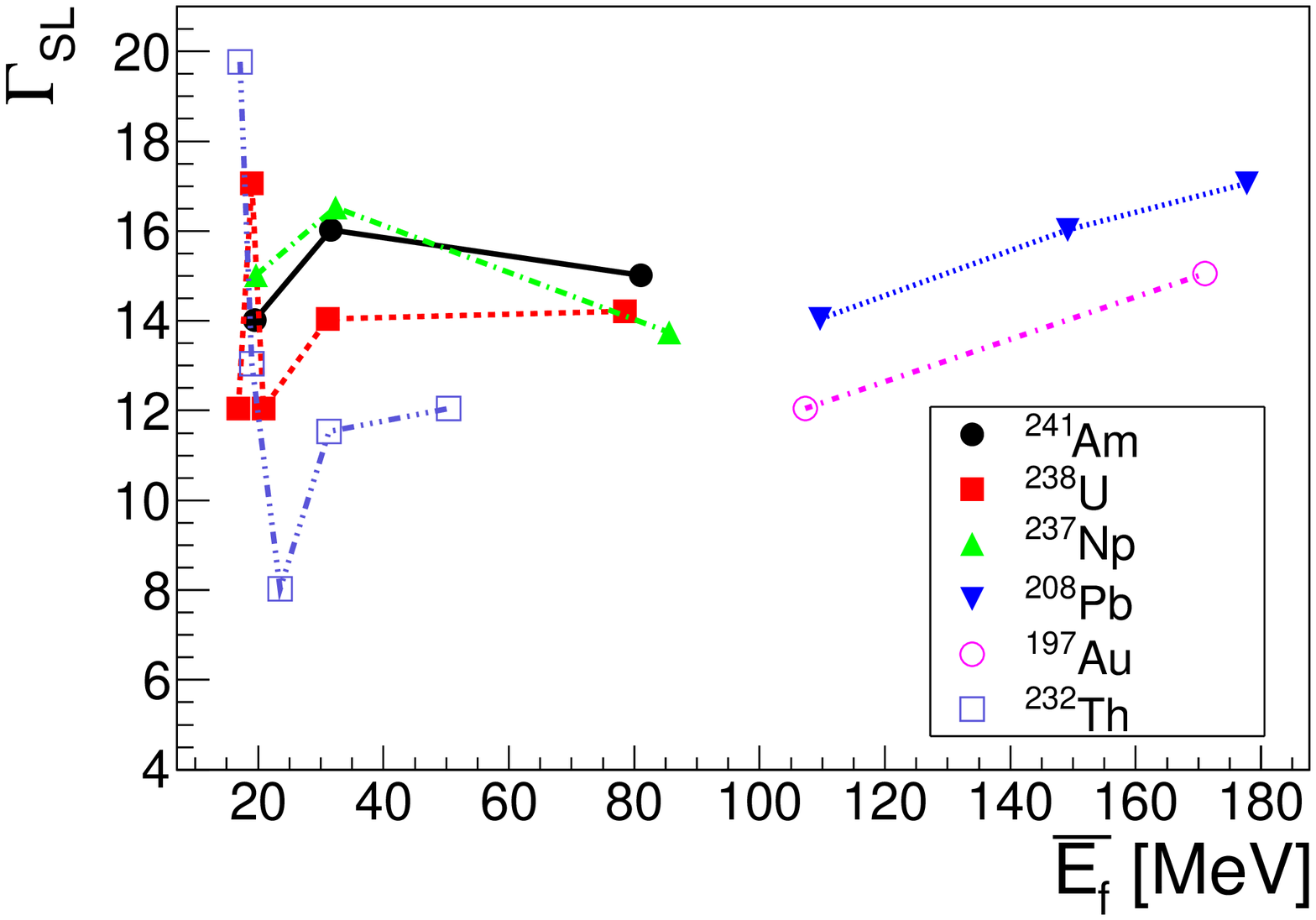}
 }
 \subfigure[Standard 1]{
    \includegraphics[scale=0.34,keepaspectratio=true]{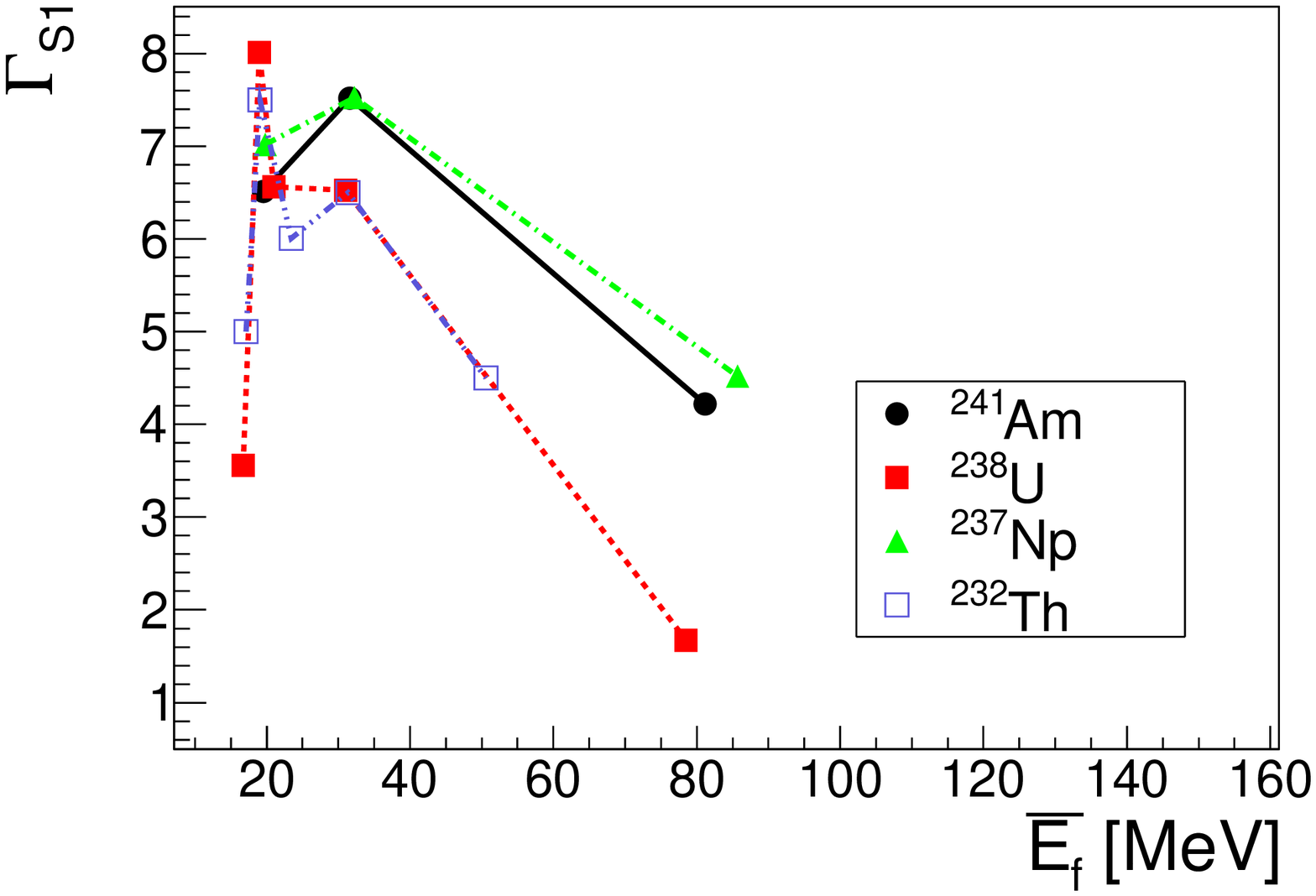}
 }
 \subfigure[Standard 2]{
    \includegraphics[scale=0.34,keepaspectratio=true]{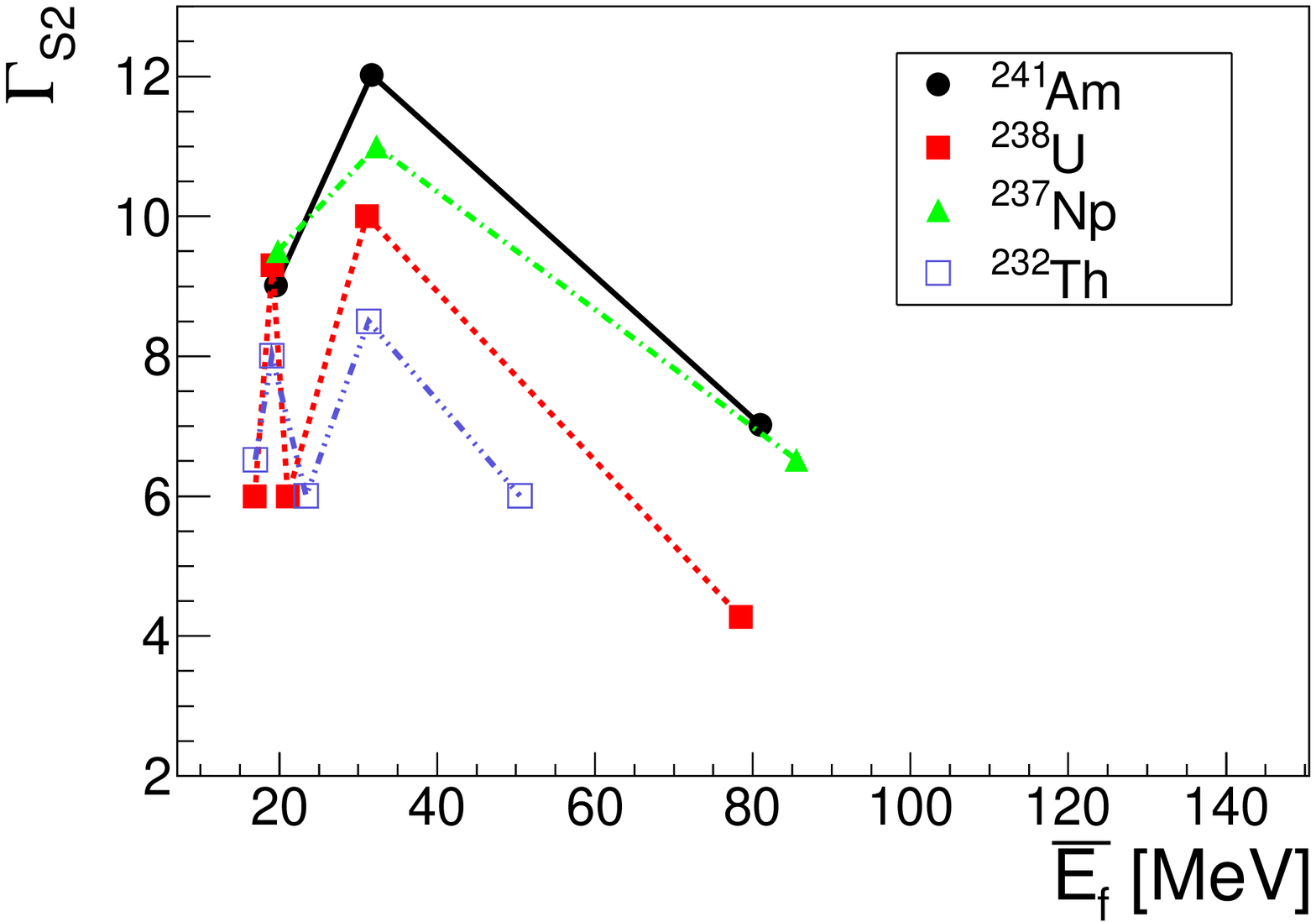}
 }
 \caption{(Color online) Behavior of the width $\Gamma_i$ for the three fission modes according to the average excitation energy $\overline{E_f}$ of the 
fissioning system. a) Superlong b) Standard 1 and c) Standard 2. Fission modes widths present an uncertainty of 0.3 mass units.}
 \label{figBehaG}
\end{figure}

\begin{figure}
 \centering
 \subfigure[Standard 1]{
    \includegraphics[scale=0.34,keepaspectratio=true]{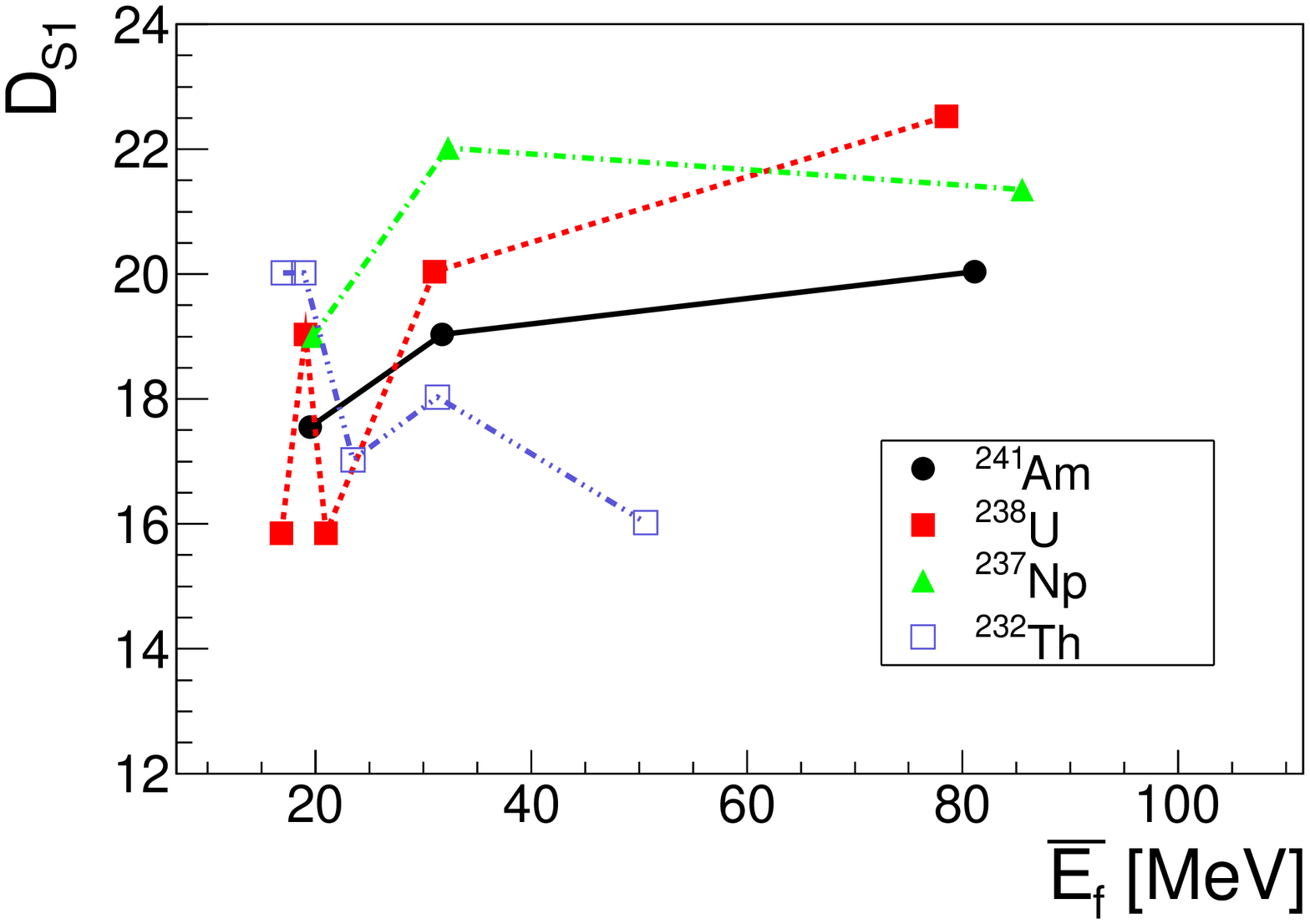}
 }
 \subfigure[Standard 2]{
    \includegraphics[scale=0.34,keepaspectratio=true]{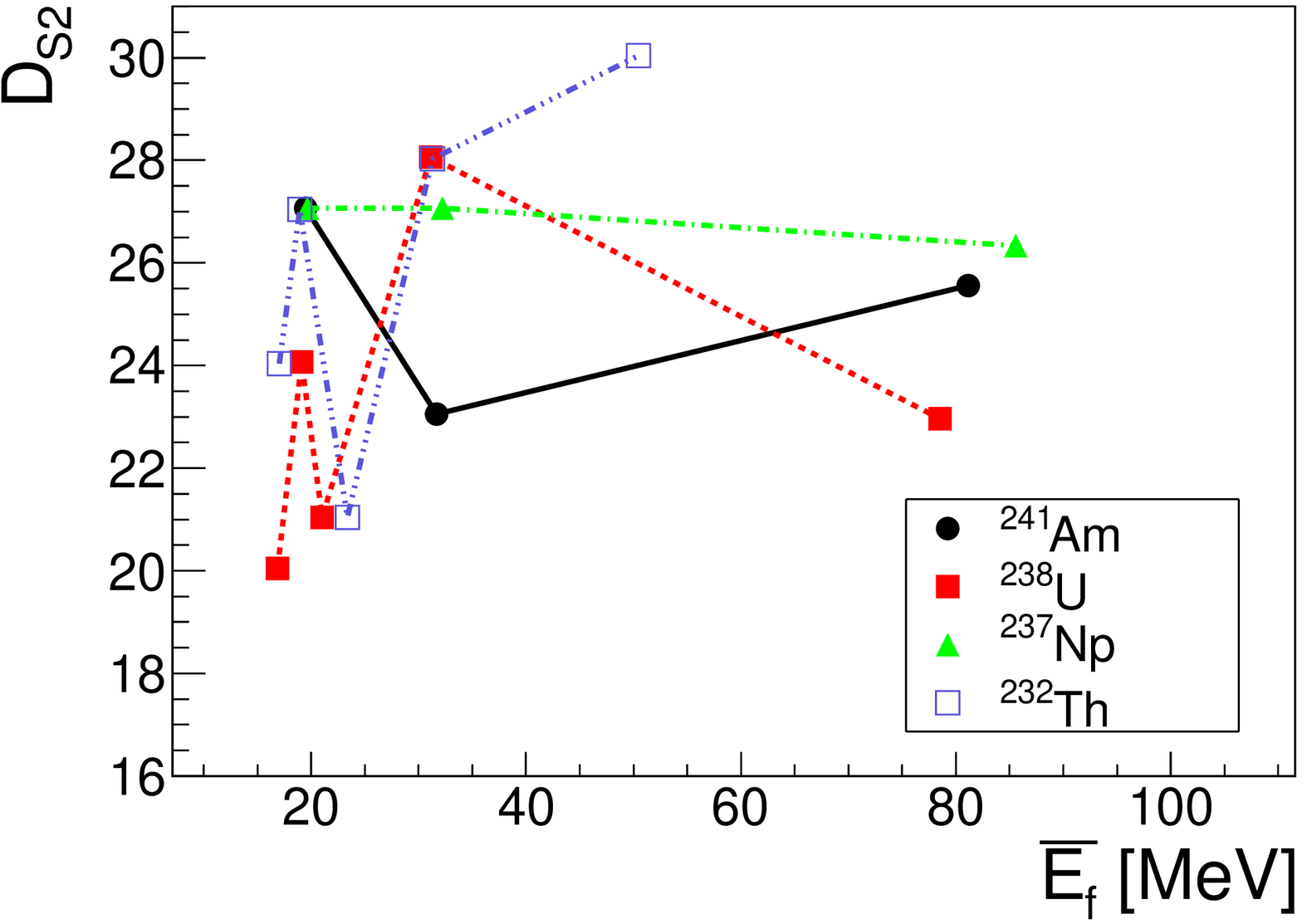}
 }
 \caption{(Color online) Behavior of the position parameter $D_i$ for the asymmetrical fission modes according to the average excitation energy $\overline{E_f}$ of the 
fissioning system. a) Standard 1 and b) Standard 2. Fission modes position parameters have an uncertainty of 0.3 mass units.}
 \label{figBehaD}
\end{figure}

Figure \ref{figBehaG} presents the behavior of widths $\Gamma_{\rm SL}$ and $\Gamma_{\rm S1,S2}$ in Equation (\ref{eqYield}). Observing the S1 and S2 width pictures together, it seems they experience some increase up to 40 MeV 
followed by a decrease that passes 80 MeV until they are no longer necessary to fit the experimental data since the reactions with $^{197}$Au and $^{208}$Pb were described only symmetrically. 
In the same region, $\Gamma_{\rm SL}$ begins with intense variation and tends to the common value of 14 mass units rapidly decreasing the 
dispersion nearly 80 MeV. After the point where there are no asymmetric fission modes, the width of the symmetrical fission takes 
on an increasing tendency. The behavior of the positions, $D_{S1}$ and $D_{S2}$, for the asymmetric fission modes in Equation (\ref{eqYield}) is presented in Figure \ref{figBehaD}.

An analysis of the details of each target and reaction in Figures \ref{figBehaK}-\ref{figBehaD} shows that the fission mode parameters of Equation (\ref{eqYield}) 
don't present a clear dependence on the fissioning system excitation energy alone.

Some final considerations regarding fission modes are necessary concerning the super-asymmetric fission mode present at reactions $^{241}$Am + 660 MeV p, 
$^{238}$U + 660 MeV p and $^{237}$Np + 660 MeV p. Table \ref{tabParSuperAsym} shows probability, width and position parameters along with the relative 
contributions of all fission modes.

\begin{table}
\centering
 \caption{Probability $P_{S3}$, width $\Gamma_{S3}$ and position $D_{S3}$ parameters for the super-asymmetric fission mode (S3) and relative contributions of all fission modes.}
\label{tabParSuperAsym}
\begin{ruledtabular}
\begin{tabular}{l  c  c  c}
			&			&			&			\\
 Parameter		& $^{241}$Am		& $^{238}$U		& $^{237}$Np		\\ 
			&			&			&			\\ \hline
$P_{S3}$		& 0.010 $\pm$ 0.002	& 0.007 $\pm$ 0.001	& 0.012 $\pm$ 0.002	\\
$\Gamma_{S3}$		& 6.5 $\pm$ 0.3		& 6.5 $\pm$ 0.3		& 7.0 $\pm$ 0.3		\\
$D_{S3}$		& 59.0 $\pm$ 0.3	& 60.0 $\pm$ 0.3	& 65.0 $\pm$ 0.3	\\\hline
&&&\\ 
\multicolumn{4}{c}{Relative contribution of the fission modes (\%)} \\
&&&\\ \hline
Standard 1 (S1)		& 2.5 $\pm$ 0.2		& 5.3 $\pm$ 0.2		& 3.0 $\pm$ 0.2		\\
Standard 2 (S2)		& 12.7 $\pm$ 0.2	& 48.1 $\pm$ 0.2	& 15.4 $\pm$ 0.2	\\
Standard 3 (S3)		& 2.6 $\pm$ 0.2		& 1.3 $\pm$ 0.2		& 2.4 $\pm$ 0.2		\\
Superlong (SL)		& 82.2 $\pm$ 0.2	& 45.3 $\pm$ 0.2	& 79.2 $\pm$ 0.2	\\
\end{tabular}
\end{ruledtabular}
\end{table}

Given the uncertainty of 0.3 mass units for both width and position parameters in this study, comparison shows that all the parameters, $P_{S3}$, $\Gamma_{S3}$ and 
$D_{S3}$, are similar for the three targets. This is only a consequence of the experimental fragments mass distributions since the values of these parameters 
are drawn exclusively from data. The relative contributions also presented in Table \ref{tabParSuperAsym} reveal that the super-asymmetric fission mode follows the 
pattern verified for the other fission channels. The relative contributions of the different fission modes in the cases of $^{241}$Am and $^{237}$Np are very similar but differ from what is obtained for the $^{238}$U case.

\section{Analysis of Nuclear Structure Influence} \label{sec:NucStrucInf}

\subsection{Critical Fissility Parameter} \label{subsec:CritFissPar}

Considering the actinides, one can see from Figures \ref{fig26MeV}-\ref{fig660_190MeV} 
that the shape of the experimental mass-yields for $^{241}$Am and $^{237}$Np targets are similar but slightly differ from $^{238}$U target where asymmetric
fission is more pronounced. In the case of $^{232}$Th, the distribution begins with asymmetrical dominance, like the other cases, 
but the symmetric fission contribution rises quickly up to 190 MeV. It is worth noting that $^{241}$Am and $^{237}$Np will show a clear symmetric contribution 
only for the 660 MeV data.

According to Chung et al \cite{Chung1981, Chung1982} the contributions of symmetric and asymmetric fission modes are closely related to the fissility parameter. Still 
according to Chung et al \cite{Chung1981, Chung1982}, although symmetric fission increases with excitation energy, the relevant mechanism would be determined by pre-scission nucleon evaporation opening new fission channels, thus modifying the nuclear structure. Therefore, the fissility parameter would  
give a more clear understanding of fission. 

A systematization of symmetric and asymmetric fission cross sections in a wide range of nuclei 
collected in \cite{Chung1981, Chung1982} suggested that it is possible to use an empirical expression to estimate the probability of different fission 
modes. In order to characterize this factor quantitatively, Chung et al \cite{Chung1981, Chung1982} introduced a critical value of the fissility 
parameter in the following form:
\begin{align}
(Z_{f}^{2}/A_f)_{cr.}=35.5+0.4(Z_f-90),
\label{crit}
\end{align}
\noindent where $Z_f$ and $A_f$ are the charge and mass number of the fissioning nucleus.

According to the critical line approach, for reactions with a higher population of fissioning systems located above the critical value of Equation 
(\ref{crit}), the symmetric fission mode should be dominant. The predominance of fissioning systems below the critical fissility parameter 
would imply an asymmetric fragment distribution. The larger contribution of symmetric fission for $^{241}$Am and $^{237}$Np than for $^{238}$U at the same energy of 
660 MeV, for instance, might be explained by a larger contribution of fissioning systems above the critical fissility line. In addition, it has been observed in another 
study \cite{Rubchenya2007} that the shell effects of the fissioning nucleus and fission fragments don't vanish completely with the increase of nuclear temperature.

From our experimental measurements, the mean mass and charge of the fissioning nuclei at 660 MeV proton energy can be calculated. The results are presented in Table \ref{tabFissParExp} where it is possible to see that the difference between $Z_f^{2}/A_f$ and $(Z_f^{2}/A_f)_{cr.}$ 
is greater for $^{241}$Am and $^{237}$Np nuclei. Therefore, the fact that symmetric fission is dominant for these nuclides but not for $^{238}$U is in 
agreement with the critical fissility parameter concept.

\begin{table}
\centering
 \caption{Charge and mass number of the fissioning system and fissility parameter calculated for $^{241}$Am, $^{238}$U and $^{237}$Np at 660 MeV from the experimental data. The critical fissility parameter is also presented for each case.}
\label{tabFissParExp}
\begin{ruledtabular}
\begin{tabular}{l  c  c  c}
			&			&			\\
 Reactions		&	$(Z_f,A_f)$	&	$Z_f^{2}/A_f$	& $(Z_f^{2}/A_f)_{cr.}$	\\ 
			&			&			&			\\ \hline
 $^{241}$Am		& 	(95,227)		&	39.76		&	37.5		\\
&&& \\
 $^{238}$U 		& 	(92,227)		&	37.29		&	36.3		\\
&&&\\
 $^{237}$Np 		& 	(93,223.4)	&	38.72		&	36.7		\\
\end{tabular}
\end{ruledtabular}
\end{table} 

\begin{figure}
 \centering
 \subfigure[$^{241}$Am + 660 MeV p]{
    \includegraphics[scale=0.35,keepaspectratio=true]{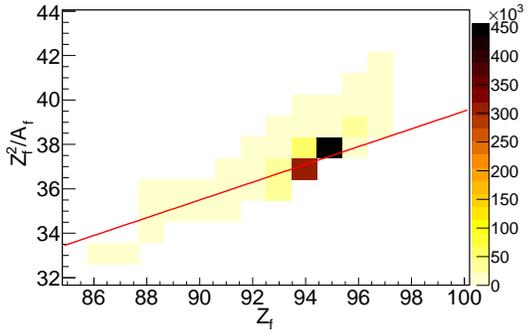}
    \label{figFissParDist1:Am}
 }
 \subfigure[$^{237}$Np + 660 MeV p]{
    \includegraphics[scale=0.35,keepaspectratio=true]{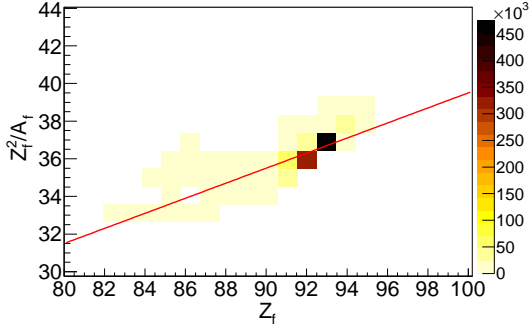}
    \label{figFissParDist1:Np}
 }
 \subfigure[$^{238}$U + 660 MeV p]{
    \includegraphics[scale=0.35,keepaspectratio=true]{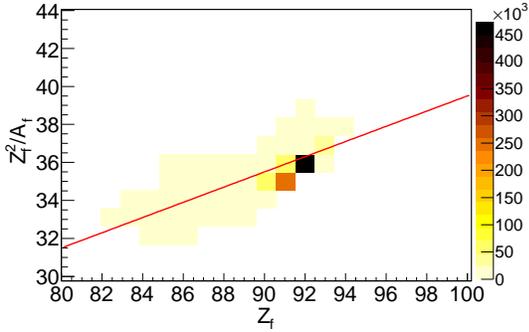}
    \label{figFissParDist1:U}
 }
 \subfigure[$^{232}$Th + 190 MeV p]{
    \includegraphics[scale=0.35,keepaspectratio=true]{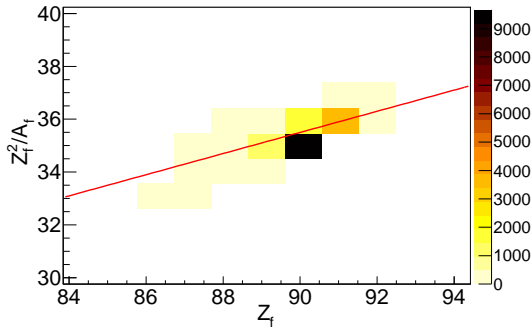}
    \label{figFissParDist1:Th}
 }
 \caption{(Color online) Distribution of the fissility parameter $Z_f^2/A_f$ for the fissioning systems for the reactions a) $^{241}$Am + 660 MeV p 
b) $^{237}$Np + 660 MeV p c) $^{238}$U + 660 MeV p and d) $^{232}$Th + 190 MeV p. The solid red line corresponds to the critical fissility parameter 
as a function of the $Z_f$ \cite{Chung1981,Chung1982}.}
 \label{figFissParDist1}
\end{figure}

\begin{figure}
 \centering
 \subfigure[$^{208}$Pb + 1000 MeV p]{
    \includegraphics[scale=0.35,keepaspectratio=true]{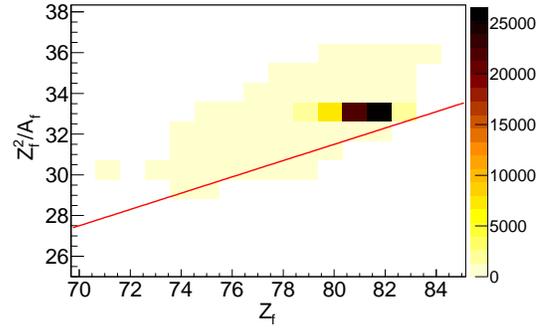}
 }
 \subfigure[$^{197}$Au + 800 MeV p]{
    \includegraphics[scale=0.35,keepaspectratio=true]{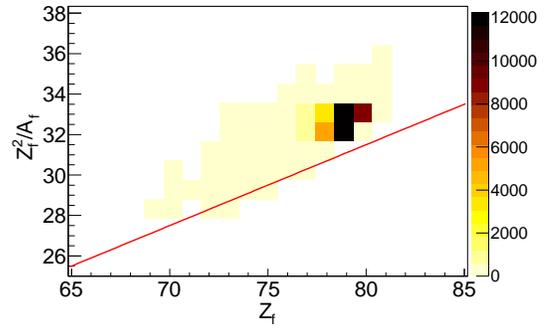}
 }
 \caption{(Color online) Distribution of the fissility parameter $Z_f^2/A_f$ for the fissioning systems for the reactions a) $^{208}$Pb + 1000 MeV p 
and b) $^{197}$Au + 800 MeV p. The solid red line corresponds to the critical fissility parameter as a function of the $Z_f$ \cite{Chung1981,Chung1982}.}
 \label{figFissParDist2}
\end{figure}

The contribution of fissioning systems above and below the critical fissility parameter can be better analyzed with the help of the CRISP model. 
In Figures \ref{figFissParDist1} and \ref{figFissParDist2} we present the distribution of the parameter $Z_f^2/A_f$ for the fissioning systems for all nuclei 
studied at the reactions with the highest energies involved for each one. Also, we calculated the ratio $R_{ab}$ between the number of fissioning nuclei located 
above the critical value and below for all reactions studied. The results are shown in Table \ref{tabRatios}, where we observe that indeed the ratio for $^{238}$U 
is much smaller than for $^{241}$Am and $^{237}$Np. These results, therefore, are in qualitative agreement with the assumption of a critical line in the 
space $Z_f^2/A_f \,\, \times \,\, Z_f$, separating the fissioning systems that will lead to symmetric or asymmetric fission \cite{Chung1981,Chung1982}.	

\begin{table}
\centering
 \caption{Ratio $R_{ab}$ between the number of fissioning nuclei located above the critical fissility value and below and ratio $R_{S,AS}$ of 
the Superlong fission mode contribution to the Standard fission modes.}
\label{tabRatios}
\begin{ruledtabular}
\begin{tabular}{l  c  c}
			&		&		\\
 Reactions		& $R_{ab}$	& $R_{S,AS}$	\\ 
			&		&		\\ \hline
 $^{241}$Am 26.5 MeV p	& 5.84		&	1	\\
 $^{241}$Am 62.9 MeV p	& 2.04		&	1.15\\	
 $^{241}$Am 660 MeV p		& 2.15		&	5.59\\
&& \\
 $^{238}$U 50 MeV (bremss)	& 0		&	0.09	\\
 $^{238}$U 26.5 MeV p		& 0		&	0.67\\
 $^{238}$U 3500 MeV (bremss)	& 0		&	0.31\\
 $^{238}$U 62.9 MeV p		& 0		&	0.95\\
 $^{238}$U 660 MeV p		& 0.07		&	0.83\\
&&\\
 $^{237}$Np 26.5 MeV p	& 4.99		&	1	\\
 $^{237}$Np 62.9 MeV p	& 0.72		&	1.43\\
 $^{237}$Np 660 MeV p		& 0.55		&	3.80\\
&&\\
 $^{232}$Th 50 MeV (bremss)	& 0		&	0.27	\\
 $^{232}$Th 26.5 MeV p	& 0.01		&	0.58	\\
 $^{232}$Th 3500 MeV (bremss)	& 0.02		&	0.48	\\
 $^{232}$Th 62.9 MeV p	& 0.05		&	0.94	\\
 $^{232}$Th 190 MeV p		& 0.19		&	3.83	\\
&&\\
 $^{208}$Pb 190 MeV p		& 2149.78	&	\texttwelveudash	\\
 $^{208}$Pb 500 MeV p		& 2229.43	&	\texttwelveudash\\
 $^{208}$Pb 1000 MeV p	& 2081.84	&	\texttwelveudash\\
&&\\
 $^{197}$Au 190 MeV p		& $>$10332	&	\texttwelveudash	\\
 $^{197}$Au 800 MeV p		& 11264.25	&	\texttwelveudash	\\
\end{tabular}
\end{ruledtabular}
\end{table} 

By comparing Figures \ref{figFissParDist1:Am}-\ref{figFissParDist1:Np} with Figures \ref{figFissParDist1:U}-\ref{figFissParDist1:Th} and recalling the results 
on Figures \ref{fig26MeV}-\ref{fig660_190MeV}, we see the indication of a connection between the critical fissility parameter and the fission dynamics.
The results on Figure \ref{figFissParDist2} also corroborates the fragment mass distributions on Figure \ref{figPbAu}. 

We can perform a different check of validity of the critical line for the fission parameter by comparing the ratio $R_{ab}$ of the fissioning systems above and 
below the critical line with the ratio 

\begin{align}
 R_{S,AS}=\frac{K_S}{2(K_{S1}+K_{S2}+K_{S3})},
\end{align}

\noindent which represents the relative contributions between symmetric and asymmetric fission modes obtained in this systematic analysis with the Random Neck Rupture Model. 
The results for all reactions are also presented in Table \ref{tabRatios}. It is possible to observe that the ratios calculated in each approach 
are in clear disagreement. Although Figures \ref{figFissParDist1} and \ref{figFissParDist2} appear qualitatively correct, this result 
shows that the calculations with the CRISP model do not confirm qualitatively the hypothesis proposed in Refs. \cite{Chung1981,Chung1982} 
for the critical fissility line. It must be emphasized, however, that the ratio $R_{ab}$ is very sensitive to the critical value, $Z_f^2/A_f$, so small variations of the critical line may result in large variations of $R_{ab}$.


\subsection{Residual nuclei and fissioning systems mass distributions} 

\begin{figure}
 \centering
 \subfigure[$^{241}$Am]{
    \includegraphics[scale=0.35,keepaspectratio=true]{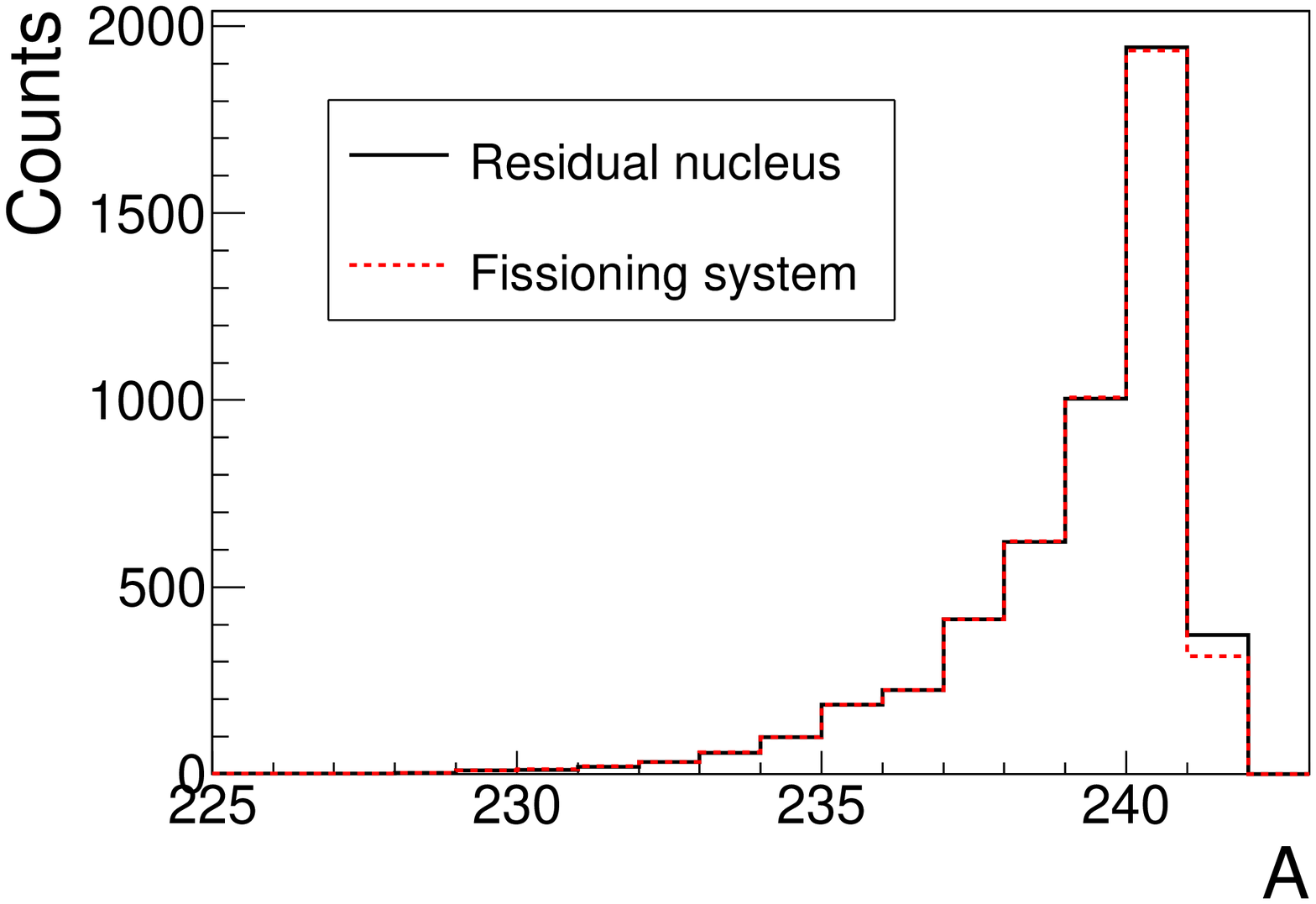}
 }
 \subfigure[$^{237}$Np]{
    \includegraphics[scale=0.35,keepaspectratio=true]{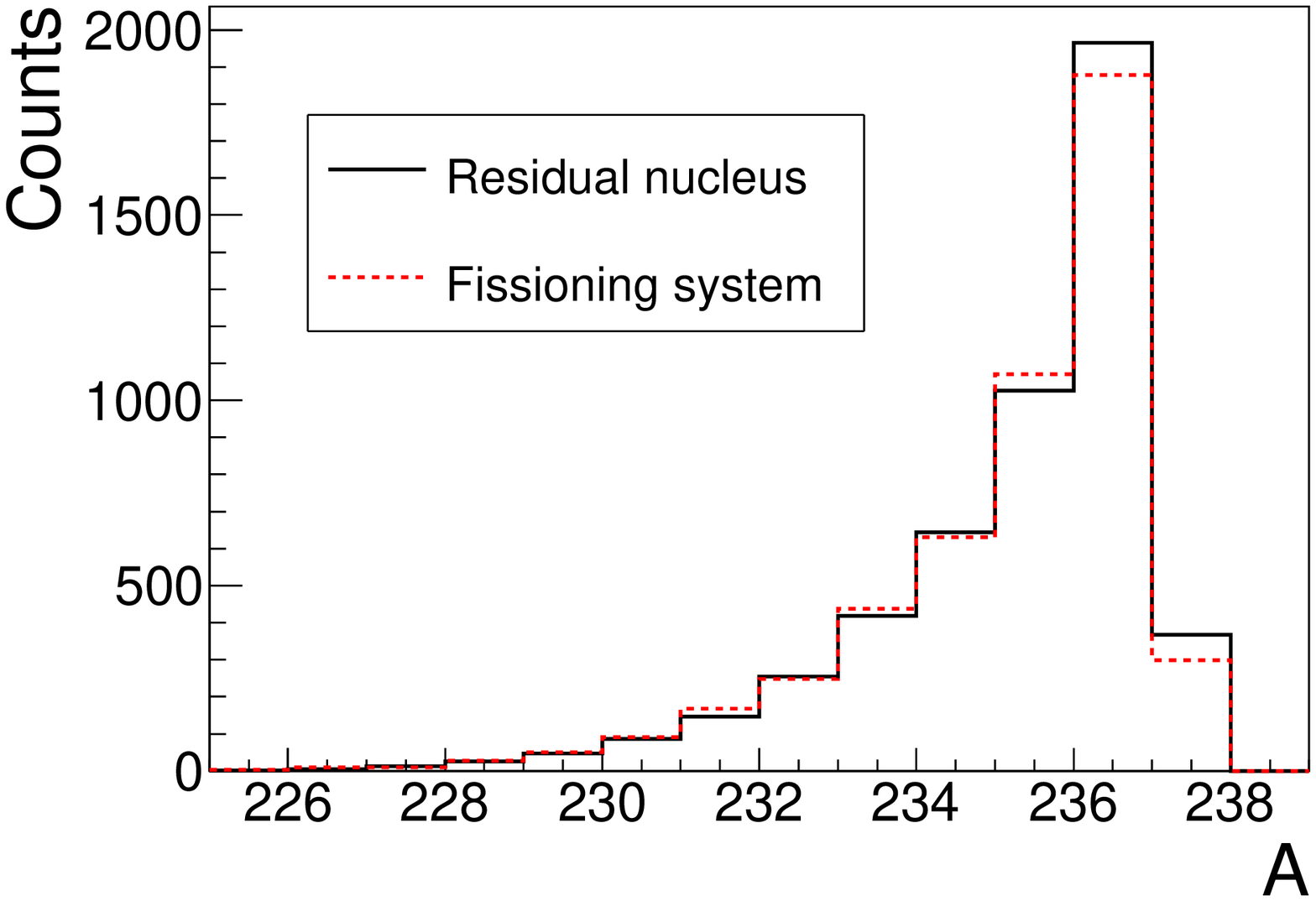}
 }
 \subfigure[$^{238}$U]{
    \includegraphics[scale=0.35,keepaspectratio=true]{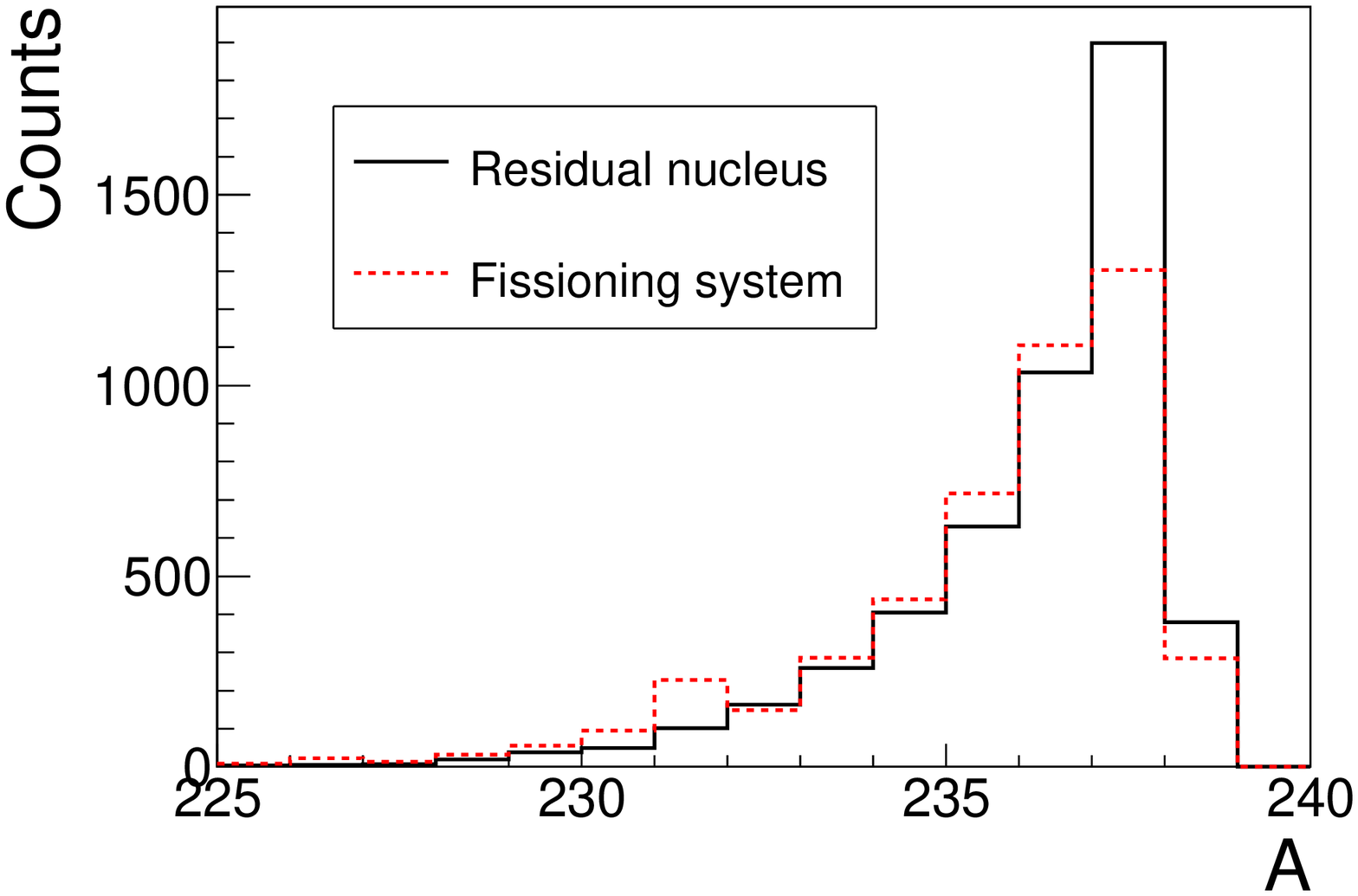}
 }
 \subfigure[$^{232}$Th]{
    \includegraphics[scale=0.35,keepaspectratio=true]{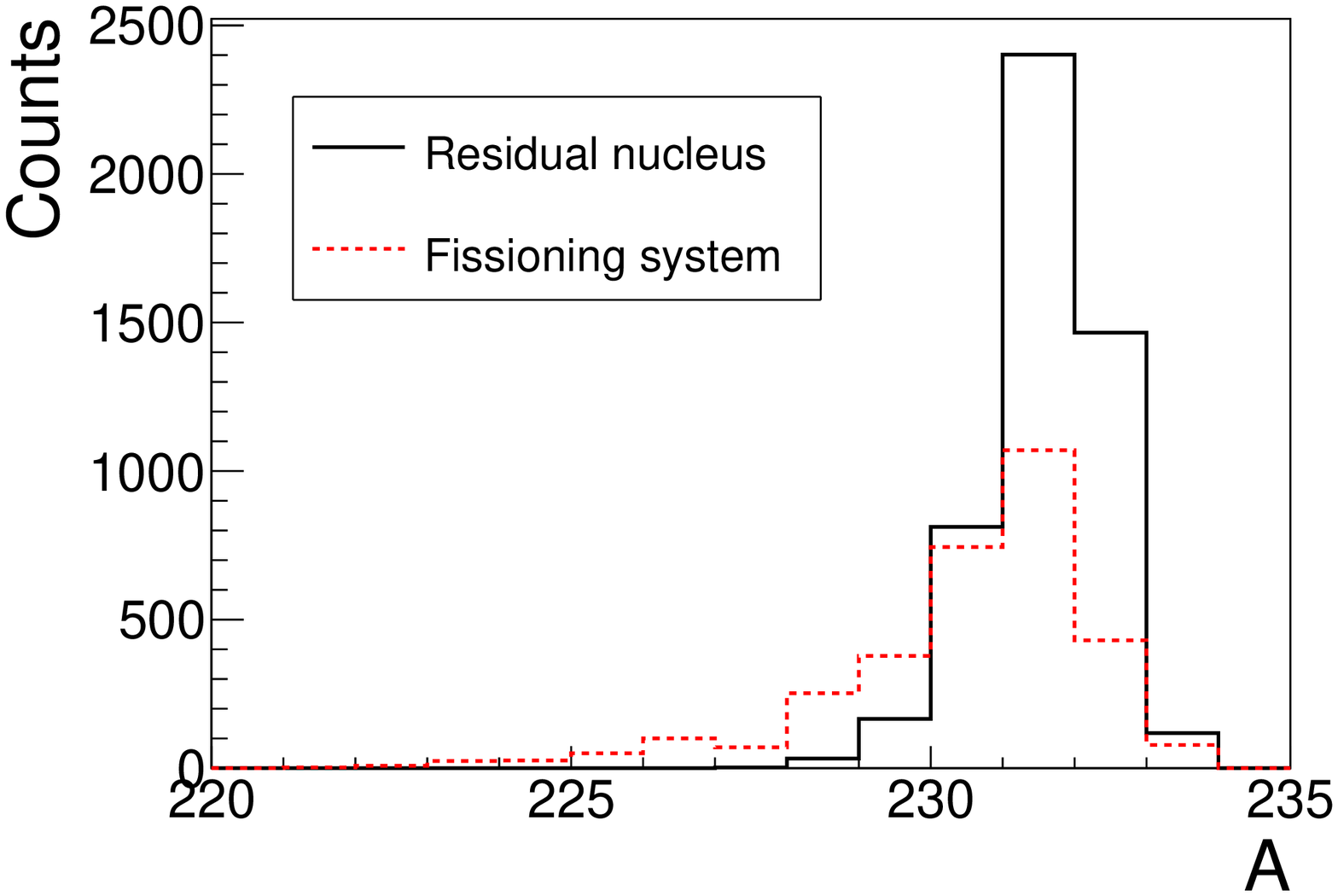}
    \label{figMassCascFissTh}
 }
\caption{(Color Online) Calculated mass distributions of the residual nuclei formed at the end of the intranuclear cascade and those of the fissioning systems for 
$^{241}$Am, $^{237}$Np and $^{238}$U reactions at 660 MeV and $^{232}$Th at 190 MeV.}
\label{figMassCascFiss}
\end{figure}

In Figure \ref{figMassCascFiss} the calculated mass distribution of the residual nuclei and that of the fissioning systems are compared for $^{241}$Am, $^{237}$Np, $^{238}$U and $^{232}$Th, as obtained with the CRISP model. We observe that indeed in the $^{238}$U case the mass distribution of the fissioning systems is rather different from the distribution of the residual nuclei, while in the cases of $^{241}$Am and $^{237}$Np their differences are not so pronounced. This behavior results from the fact that $^{238}$U has a smaller fissility when compared to 
$^{237}$Np and $^{241}$Am, and therefore the evaporation of nucleons is more relevant in that case. But it should be noticed that the exact same reasoning 
holds for $^{232}$Th which has a fissility even smaller than for $^{238}$U. Figure \ref{figMassCascFissTh} shows that in the case of $^{232}$Th the mass distributions of the residual nuclei and of the fissioning systems differ even more from each other. 

One can observe that the evaporation process, which is longer for $^{238}$U and $^{232}$Th than for $^{241}$Am and $^{237}$Np, play a determinant role in the definition of 
the asymmetric fission contribution.


\subsection{Neutron excess}

From a more careful analysis of the reactions on $^{232}$Th, one can observe that, although important, the critical fission parameter cannot describe all features of the symmetric and asymmetric fission modes. Indeed, in Figure \ref{figFissParDist1} we observe that the distributions for $^{238}$U and $^{232}$Th are very similar, both of them being mostly below the critical line, while for $^{241}$Am and $^{237}$Np most of the fissioning systems lie above the critical line. However, in Figures \ref{fig26MeV}-\ref{fig660_190MeV} we notice that the contribution of symmetric fission is more relevant in the case of $^{232}$Th than in the case of $^{238}$U, in 
contradiction with the ratio $R_{ab}$ in Table \ref{tabRatios}.

Yet another way of examining the relation 
between charge and mass number that can complement the analysis of the fissility parameter is by studying the neutron excess. 

\begin{figure}
 \centering
 \includegraphics[scale=0.43,keepaspectratio=true]{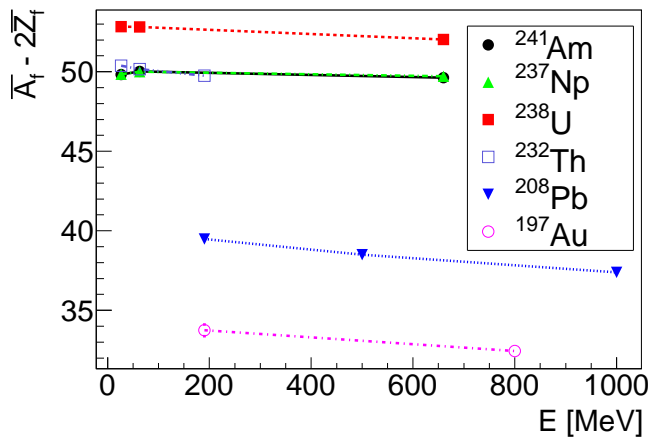}
 \caption{(Color Online) Average neutron excess ($\overline{A_f} - 2\overline{Z_f}$) of the fissioning systems for the all reactions in this study as function of the interaction energy.}
 \label{figNeuExces}
\end{figure}

The average neutron excess of the fissioning systems for all reactions in this study as a function of the interaction energy is shown in Figure \ref{figNeuExces}. 
It is well-known that the fissility of these nuclei decrease linearly with decreasing charge number and so does the fissility parameter. The neutron excess, on 
the other hand presents a different behavior. $^{238}$U shows the highest neutron excess in all energies staying above 52 neutrons. $^{241}$Am and $^{237}$Np present identical behavior regarding neutron excess staying around 50 neutrons. This region of neutron excess is also occupied by $^{232}$Th. $^{208}$Pb and $^{197}$Au have the lower neutron excess among the nuclei in this study. The order in which all nuclei are grouped in terms of decreasing neutron excess seems to reproduce the order in which symmetric fission grows dominant. Table \ref{tabNeutExce} details the information on this subject.

\begin{table}
\centering
 \caption{Average neutron excess of the fissioning systems for all proton-induced reactions studied.}
\label{tabNeutExce}
\begin{ruledtabular}
\begin{tabular}{l  c }
			&			\\
 Reactions		& $\overline{A_f} - 2\overline{Z_f}$	\\ 
			&			\\ \hline
 $^{241}$Am 26.5 MeV p	& 49.820 $\pm$ 0.001		\\
 $^{241}$Am 62.9 MeV p	& 50.020 $\pm$ 0.002		\\
 $^{241}$Am 660 MeV p	& 49.620 $\pm$ 0.004	\\
& \\
 $^{238}$U 26.5 MeV p	& 52.84	$\pm$	0.01	\\
 $^{238}$U 62.9 MeV p	& 52.82	$\pm$	0.01	\\
 $^{238}$U 660 MeV p		& 52.02	$\pm$	0.01	\\
&\\
 $^{237}$Np 26.5 MeV p	& 49.84	$\pm$	0.01	\\
 $^{237}$Np 62.9 MeV p	& 50.01	$\pm$	0.01	\\
 $^{237}$Np 660 MeV p	& 49.72	$\pm$	0.01	\\
&\\
 $^{232}$Th 26.5 MeV p	& 50.38	$\pm$	0.01	\\
 $^{232}$Th 62.9 MeV p	& 50.16	$\pm$	0.01	\\
 $^{232}$Th 190 MeV p	& 49.76	$\pm$	0.02	\\
&\\
 $^{208}$Pb 190 MeV p	& 39.48	$\pm$	0.18	\\
 $^{208}$Pb 500 MeV p	& 38.50	$\pm$	0.08	\\
 $^{208}$Pb 1000 MeV p	& 37.38	$\pm$	0.06	\\
&\\
 $^{197}$Au 190 MeV p	& 33.74	$\pm$	0.38	\\
 $^{197}$Au 800 MeV p	& 32.44	$\pm$	0.09	\\
\end{tabular}
\end{ruledtabular}
\end{table}

These observations seem to confirm the hypothesis that neutron deficiency plays some role in the determination of symmetric or asymmetric fission  \cite{Duijvestijn1999,Duijvestijn2001}. The study of both neutron excess and the critical fissility line could help providing a more detailed description 
of the mechanism that leads to symmetry and asymmetry in fission.

\section{Cold Cluster Emission} \label{sec:ColdClus}

One can notice that in the case of the reaction $^{197}$Au + 800 MeV p there is a shoulder 
on the left side of the fragment mass distribution (lighter fragments) with no similar structure on the right side (heavier fragments). 
One possible explanation is the existence of a process that resembles a super-asymmetric fission mode but that is, in fact, 
a cold cluster emission. This process would leave all of the system excitation energy for the heavy fragment which would evaporate at a rate above the 
average and by this way smoothing the hump on the right side of the distribution.

\begin{figure}
 \centering
 \includegraphics[scale=0.35,keepaspectratio=true]{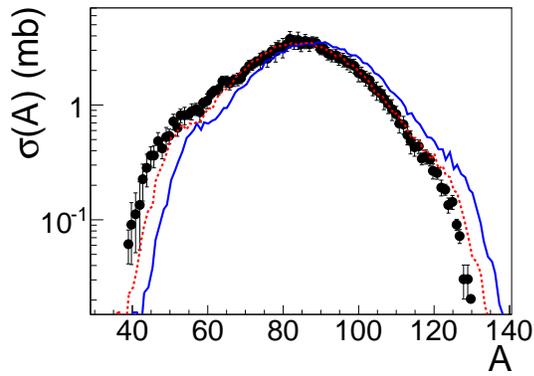}
 \caption{(Color Online) Fragment mass distribution for the reaction $^{197}$Au + 800 MeV proton for which the cold emission hypothesis was tested (blue solid line). 
The red dotted line represent the same calculation shifted by an 4 mass units. Experimental data from \cite{Benlliure2001}.}
 \label{figAu197_ColdEmi}
\end{figure}

In Figure \ref{figAu197_ColdEmi}, the results of the CRISP model after the implementation of the hypothesis just described are presented. We observe a 
disagreement between calculation and data but the general trend of the distributions are similar and the calculated one seems to be shifted to higher masses 
by nearly 4 mass units, reflecting a lack of nucleon emission in the calculation. The same calculation but corrected to the left by 4 mass units is presented 
also in Figure \ref{figAu197_ColdEmi}. This reinforces the similarities between shapes.

The missing nucleons cannot be from post-scission emission since the light cluster is supposed to be cold, therefore the left-side hump in the distribution 
would not move under this process. Pre-scission emission can happen in the intranuclear cascade and during the evaporation/fission competition process. 
Modification in our model to correct the lack of nucleon emission must be carried out through a complete analysis of fission, spallation and spectrum of 
emitted neutrons and protons. This work is under development and will be presented in a forthcoming paper. However, the fact that the shape of the distribution 
is correct is an evidence of the emission of cold clusters. 

A hypothesis by Moretto \cite{Moretto1975} states that fission and evaporation of clusters could be 
different aspects of a general process called binary emission. In this way a common description for evaporation and fission would be possible. Moreover, spontaneous 
fission has already been described as cluster emission \cite{Duarte2002}. Nevertheless, in the super-asymmetric fission both light and heavy fragments are excited. 
The evidence of a process in which a cold cluster is emitted can signalize that fission and cluster emission cannot be understood as a single process. 

The argument previously presented for the neutron deficiency favoring symmetric fission still holds since as far as the
fission of $^{197}$Au at 800 MeV is concerned it is only symmetric, with cold cluster emission accounting for the other structures visible in the mass distribution. 
Similar considerations can be made for the distribution in Figure \ref{figPbAu:Pb1000} where some deviation from the multi-modal fission approach can also 
be observed although to a lower extent. Of course, new experimental data 
on target nuclei of similar masses are needed in order to make definitive conclusions on this matter.


\section{Conclusions} \label{sec:Conc}

In this paper we present new experimental data on mass distribution of fission fragments from $^{241}$Am proton-induced fission at $660$ MeV measured at the 
LNR Phasotron (JINR), and a systematic analysis of several measured fragment mass distributions from different fission reactions available in the literature. 
The analysis was performed in the framework of the Random Neck Rupture Model, with the inclusion of the symmetric mode, called Superlong and two asymmetric 
modes, called Standard 1 and 2. In some cases we found necessary to include a fourth mode associated with super-asymmetric fission, called Standard 3.

From the new experimental data we have found that the fragment mass distribution for the $^{241}$Am target presents a prominent symmetrical fission signal, 
with a smaller presence of the two usual asymmetric modes.

From the systematic analysis, we conclude that the fission dynamics is more directly dependent on the fissioning system structure, not on its excitation 
energy. Also, the $Z^2_f/A_f$ criterion proposed by Chung et al \cite{Chung1981,Chung1982} qualitatively describes the experimental data but needs improvements 
from the quantitative point of view. Also, a joint analysis of the excitation energy, the critical fissility line and the neutron excess seems to provide a more complete view of the fission process. 

The super-asymmetric fission mode was observed for three targets, $^{241}$Am, $^{238}$U and $^{237}$Np, and it was found that the parameters for this channel are 
similar for all three, given the uncertainty. Also, the relative contribution of the super-asymmetric channel shows the pattern already verified for the other fission 
channels regarding the fact that $^{241}$Am and $^{237}$Np are very similar to each other but differ from $^{238}$U.

The analysis of the fission fragment mass distribution from the reaction $^{197}$Au + 800 MeV p suggests a new kind of fragment distribution which cannot be 
described by the super-asymmetric fission. We propose in this work another mechanism for the production of those fragments through the emission of cold 
clusters. This new mechanism indicates that fission and cluster evaporation may differ by the excitation energy in the light fragment. 

New experimental results from the LNR Phasotron (JINR) on the fission of Bi and Pb by 660 MeV protons will be available in the future. This will be an 
opportunity to extend the systematic analysis of this work and test the suggestions that were made.

\begin{acknowledgments}
E. Andrade-II acknowledge the support from the Brazilian federal agency CAPES. G. Karapetyan is grateful to CNPq Grant No. 112986/2015-3 and to International 
Centre for Theoretical Physics (ICTP) under the Associate Grant Scheme. We also acknowledge the support from the Brazilian agency FAPESP at S\~ao Paulo State.  
\end{acknowledgments}

\input{fission_systematics.bbl}

\end{document}